\documentclass[aps,prl,10pt,twocolumn,superscriptaddress]{revtex4-1}

\usepackage{footnote}
\usepackage{times}
\usepackage{booktabs}
\usepackage{bm,natbib}
\usepackage{inputenc}
\usepackage{graphicx}
\usepackage{mathrsfs}
\usepackage{dcolumn,fancyhdr}
\usepackage{amsmath}
\usepackage{amssymb}
\usepackage{amsfonts,gensymb}
\usepackage{indentfirst}
\usepackage{bbold}
\usepackage{multirow}
\usepackage{dsfont}
\usepackage[usenames,dvipsnames]{xcolor}      

\usepackage{changes}

\usepackage{hyperref}
\hypersetup{
colorlinks=true,
linkcolor=blue,
citecolor=blue,
filecolor=magenta,
urlcolor=cyan
}

\usepackage[capitalize]{cleveref}

\newcommand{\tr}[1]{\mathrm{Tr}\left[ {#1} \right]}

\newcommand{\bOmega}{\boldsymbol \Omega}

\begin{document}
\title{Conditional dynamics of optomechanical two-tone backaction-evading measurements}
\author{Matteo Brunelli}
\affiliation{Cavendish Laboratory, University of Cambridge, Cambridge CB3 0HE, United Kingdom}
\author{Daniel Malz}
\affiliation{Max-Planck-Institut f\"ur Quantenoptik, Hans-Kopfermann-Strasse 1, D-85748 Garching, Germany}
\author{Andreas Nunnenkamp}
\affiliation{Cavendish Laboratory, University of Cambridge, Cambridge CB3 0HE, United Kingdom}

\date{\today}

\begin{abstract}
Backaction-evading measurements of mechanical motion can achieve precision below the zero-point uncertainty and quantum squeezing, which makes them a resource for quantum metrology and quantum information processing. We provide an exact expression for the conditional state of an optomechanical system in a two-tone backaction-evading measurement beyond the standard adiabatic approximation and perform extensive numerical simulations to go beyond the usual rotating-wave approximation. We predict the simultaneous presence of conditional mechanical squeezing, intra-cavity squeezing, and optomechanical entanglement. We further apply an analogous analysis to the multimode optomechanical system of two mechanical and one cavity mode and find conditional mechanical Einstein-Podolski-Rosen entanglement and genuinely tripartite optomechanical entanglement. Our analysis is of direct relevance for ultra-sensitive measurements and measurement-based control in high-cooperativity optomechanical sensors operating beyond the adiabatic limit.
\end{abstract} 
 
\maketitle

\emph{Introduction.}---The standard quantum limit (SQL) is the precision limit that arises from the fundamental trade-off between the information extractable from a measurement and the associated backaction when continuously monitoring the mechanical motion \cite{QuantumNoiseRev, OnofrioRev}. Backaction-evading (BAE) measurements bypass this limit by restricting the measurement to a single quadrature of motion \cite{Braginsky1, QNDThorne78, WeakMeasRev}. One way to implement this is to parametrically couple the mechanical motion to a cavity driven on both mechanical sidebands \cite{Braginsky1, CMJ2008}. BAE measurements have been demonstrated in optomechanics, 
with sensitivities approaching the SQL \cite{BAEOpto09, BAEOpto14, DanielExp18}, and 
exploited to generate spin squeezing in light-controlled atomic ensembles~\cite{BAESpin15}. 
They have also been extended to collective observables of two modes~\cite{TsangCohCanc, TsangQND,TwoModeBAE,TwoModeBAEExp,HammererEPR, BAEBEC,HybridBAE}.

Recent experimental advances have allowed to access the conditional dynamics and real-time feedback of weakly monitored optomechanical systems at the quantum limit~\cite{Wilson2015,Shudir2017,Rossi2017,Rossi2018}. In BAE measurements, continuous monitoring would enable uncertainties below the SQL and the generation of conditional mechanical squeezing~\cite{Wiseman93, QuantumMeasBook, JacobsSteckRev}, which is a valuable resource in quantum metrology~\cite{NonClassMetrology,SqueezingMetrology} and continuous-variable quantum information~\cite{GaussRev,Oussama15,Oussama18}. Surprisingly, the current literature only considers an approximate description of this process, based on the intracavity field adiabatically following the mechanical motion~\cite{Doherty99}. With state-of-the-art cavity optomechanics experiments operating in the backaction-dominated regime \cite{QBA18}, this description has become inadequate. 

\begin{figure}[t!]
\centering
\includegraphics[width=0.9\linewidth]{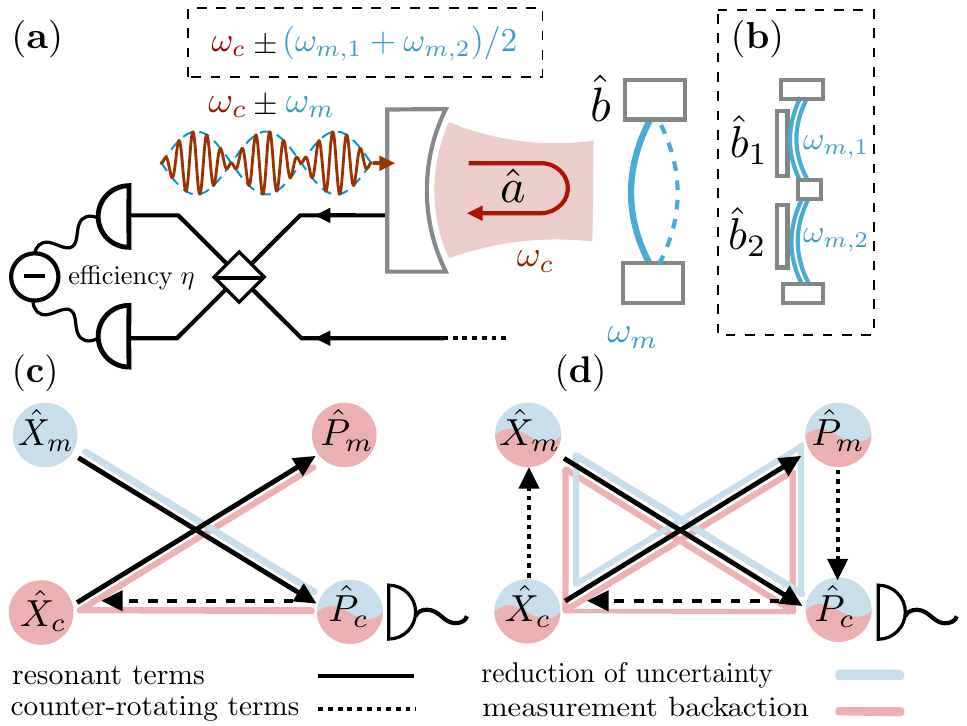}
\caption{({\bf a}) Backaction-evading (BAE) measurement of a single mechanical quadrature. An optomechanical cavity ($\hat a$) is driven on the lower and upper mechanical ($\hat b$) sideband and is continuously monitored via the output homodyne current. ({\bf b}) If two mechanical modes $\hat b_1$ and $\hat b_2$ are considered instead (dashed boxes), a two-mode BAE measurement is realized. ({\bf c}) Monitoring of the output field both introduces backaction and allows to extract information. Arrows originate from the source terms in the equations of motion derived from Eq.~\eqref{Hint}. Within RWA, backaction is confined to $\hat P_m$ and reduction of uncertainty to $\hat X_m$. ({\bf d}) Counter-rotating terms open new channels, which corrupt the BAE regime and reduce squeezing in $\hat X_m$, but at the same time enable joint reduction of uncertainty of other variables, e.g. $\hat X_c$ and $\hat P_m$; correlations are thus enhanced and robust entanglement can be generated subject to measurement.
\label{f:Sketch}}
\end{figure}

In this Letter we present an exact treatment of the conditional dynamics of BAE measurements beyond adiabatic elimination and valid for initial Gaussian states. 
We predict the existence of an optimal value of mechanical squeezing
(in terms of the system's parameters). 
We then numerically go beyond the rotating-wave approximation (RWA) and describe the quantum features induced by the measurement on the whole optomechanical system
(conditional intra-cavity squeezing and optomechanical entanglement) which are entirely missed by taking the adiabatic approximation. 

We finally extend our analysis to two mechanical modes coupled to a common cavity field. We show both conditional generation of mechanical Einstein-Podolski-Rosen (EPR) as well as genuine tripartite optomechanical entanglement. Our study provides a substantial improvement in the description of weakly monitored optomechanical systems (as well as parametrically coupled superconducting circuits~\cite{Hacohen-Gourgy2016,Chantasri2016,Murch2013}) and opens novel avenues for ultra-sensitive measurements and measurement-based quantum control of mechanical motion.

\begin{figure}[t]
\centering
\includegraphics[width=0.9\linewidth]{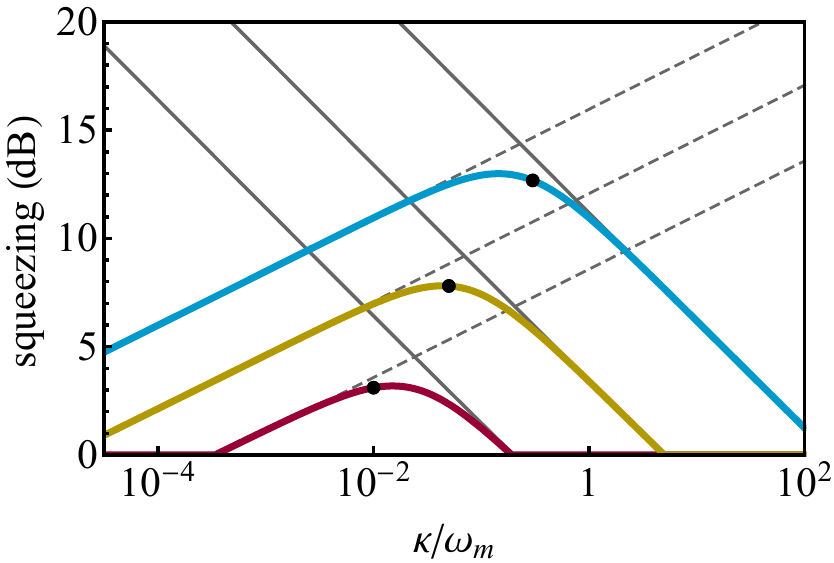}
\caption{Mechanical squeezing (in dB) for $g=0.01\omega_m$ (red), $g=0.05\omega_m$ (yellow) and $g=0.3\omega_m$ (cyan) as predicted by Eq.~\eqref{VarQcond}. Other parameters are $\gamma=10^{-4}\omega_m,\, \bar n=10,\,\eta=1$. Solid black lines represent the adiabatic solution $\sigma^2_{X_m,\mathrm{ad}}$ while dashed lines that of a slow cavity $\sigma^2_{X_m,\mathrm{slow}}$. For each curve, the part to the left of the black dot ($g=\kappa$) is in the strong-coupling regime.
\label{f:PlotSqRWA}}
\end{figure}

\emph{Optomechanical conditional dynamics.}---We consider an optomechanical system where a mechanical oscillator of frequency $\omega_m$ modulates the frequency of a cavity mode $\omega_c$~\cite{OptoRev}. The Hamiltonian is given by ($\hbar=1$)
\begin{equation}
\hat H=\omega_c \hat a^\dag \hat a+\omega_m \hat b^\dag \hat b \label{H0}- g_0 \hat a^\dag \hat a (\hat b+\hat b^\dag)+\mathcal{E}(t) \hat a^\dag+\mathcal{E}^*(t) \hat a\, ,
\end{equation}
where $\hat a$ ($\hat b$) describes the cavity (mechanical) mode, $g_0$ is the single-photon coupling strength, and the cavity is driven on both mechanical sidebands $\omega_c\pm\omega_m$ with the same strength, i.e., $\nobreak{\mathcal{E}(t)=2\vert\mathcal{E}\vert e^{-i \omega_c t}\cos\omega_mt}$. After 
linearization and moving to an interaction picture with respect to the free mechanical and cavity evolution, we obtain 
\begin{equation}\label{Hint}
\hat H_I(t)=-g \hat X_c \left[\hat X_m (1+\cos2\omega_m t)+\hat P_m \sin2\omega_m t \right] \,,
\end{equation}
with coupling strength $g\equiv g_0\vert\mathcal{E}\vert/\sqrt{\omega_m^2+\kappa^2/4}$, cavity decay rate $\kappa$,
and dimensionless quadratures $\nobreak{\hat X_c=(\hat a+\hat a^\dagger)/\sqrt{2}}$, $\nobreak{\hat X_m=(\hat b+\hat b^\dagger)/\sqrt{2}}$, and $\nobreak{\hat P_m=i(\hat b^\dagger-\hat b)/\sqrt{2}}$.
Eq.~\eqref{Hint} has a time-independent part, $\hat H_{\mathrm{QND}}=- g\hat X_c \hat X_m$, and an oscillating part $\hat H_{\mathrm{CR}}(t)$.
In the good-cavity limit $\kappa\ll\omega_m$, the latter can be neglected and the interaction is manifestly QND~\cite{CMJ2008}.

We also include interactions with the photonic and the mechanical environment~\cite{SM}. 
Both environments consist of a collection of uncorrelated modes that interact with the system at time $t$ and are otherwise uncoupled;  this assumption both gives rise to a Markovian environment and provides a monitoring channel. After interacting with the system the photonic modes of the environment undergo a homodyne measurement of the {\it phase quadrature} $\hat P_c$~\cite{QuantumMeasBook} [see Fig.~\ref{f:Sketch} ({\bf a})].

Given the (bi)linear nature of both the interaction and the measurement and given a Gaussian initial state, the state of the optomechanical system $\hat \varrho$ is exhaustively described in terms of the mean vector $\bar{\rm{x}}=\tr{\hat\varrho \,\hat{\rm{x}}}$ and covariance matrix (CM) $\nobreak{\sigma=\tfrac12\tr{\hat\varrho \{ \hat{\rm{x}}-\bar{\rm{x}},(\hat{\rm{x}}-\bar{\rm{x}})^T\}}}$, where we set
$\nobreak{\hat {\rm{x}}=(\hat X_c,\hat P_c,\hat X_m, \hat P_m)^T}$~\cite{GaussRev}. The conditional evolution of the continuously monitored system is then described by the following set of equations~\cite{Marco16,AlessioBook}
\begin{align}
d \bar{\rm x} &=A \bar{\rm x} dt-(\sigma B-N) dW_t\, , \label{Cond1}\\
\dot\sigma&=A\sigma +\sigma A^T+D -(\sigma B-N)(\sigma B-N)^T\label{Cond2}\, ,
\end{align}
where $A=A(t)$ is the drift matrix, $D$ the diffusion matrix, $B$ and $N$ account for the reduction of uncertainty and added noise due to the measurement; $W_t$ is a vector of independent Wiener processes ($dW_jdW_k=\delta_{jk}dt$), see SM~\cite{SM}. Notice that the stochastic evolution, consequence of the measurement-induced disturbance, is confined to the first moments. Therefore, at any time the conditional state is represented by a Gaussian state whose CM evolves deterministically according to Eq.~\eqref{Cond2}. This will represent the main tool of our analysis.

\begin{figure*}[t]
\includegraphics[width=\linewidth]{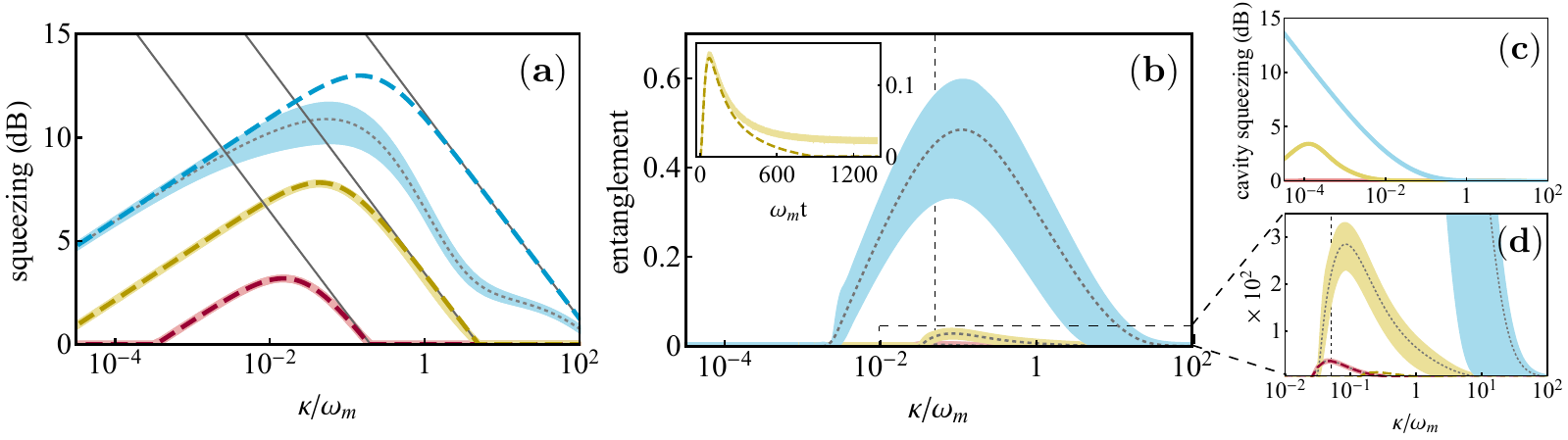}
\caption{({\bf a}) Mechanical squeezing (in dB) assuming the RWA [Eq.~\eqref{VarQcond}] (dashed dark curves) and beyond the RWA (lighter shaded areas). 
The curves are for $g=0.01\omega_m$ (red), $g=0.05\omega_m$ (yellow) and $g=0.3\omega_m$ (cyan); the dotted curve shows the mean squeezing (averaged over one mechanical period) and the shaded area extends between the minimum and maximum value of squeezing.  Solid black lines represent the adiabatic solution $\sigma^2_{X_m,\mathrm{ad}}$. ({\bf b}) Conditional entanglement (measured by the logarithmic negativity) for the same couplings as ({\bf a}); the vertical dashed line corresponds to $\kappa=0.05\omega_m$ and in the inset we show the temporal evolution of entanglement along this cut for the case $g=0.05\omega_m$. ({\bf c}) Conditional cavity squeezing for the same couplings as ({\bf a}). ({\bf d}) Zoom-in of panel ({\bf b}). In all panels other parameters are: $\gamma=10^{-4}\omega_m,\, \bar n=10,\,\eta=1$. \label{f:PlotSq}}
\end{figure*}

\emph{Mechanical squeezing beyond adiabatic approximation.}---We start by studying the conditional dynamics of a two-tone BAE measurement within the RWA, namely when Eq.~\eqref{Hint} reduces to the perfect QND interaction $\hat H_{\mathrm{QND}}=- g\hat X_c \hat X_m$. The steady-state conditional CM~\eqref{Cond2} can be obtained analytically (cf. SM \cite{SM}). Here,  we focus on the variances of the two mechanical quadratures
\begin{align}
\sigma^2_{X_m}&=\frac{\sqrt{\gamma^2+\kappa^2+2\zeta}}{16g^2\eta \kappa}
\left(\zeta+\gamma^2-\gamma\sqrt{\gamma^2+\kappa^2+2\zeta} \right), \label{VarQcond} \\
\sigma^2_{P_m}&=\bar n+\frac12+\frac{2g^2}{\gamma(\gamma+\kappa)} \label{VarPcond} \, ,
\end{align}
where $\zeta=\sqrt{\gamma\kappa[16 g^2\eta(1+2\bar n)+\gamma\kappa]}$, $\bar n$ is the thermal occupancy of the mechanical bath and $0\le \eta\le1$ is the quantum efficiency of the measurement. These exact expressions are the first central result of our work.

We note that for $\eta\rightarrow0$ no measurement is recorded and Eq.~\eqref{VarQcond} reduces to the unconditional variance $\sigma^2_{X_m} \rightarrow \bar n+\tfrac{1}{2}$, which is consistent with the fact that $\hat X_m$ is a conserved quantity. 
Physically, the presence of a monitoring channel introduces disturbance, which directly affects the conjugate quadrature ($\hat X_c$) and, via the optomechanical coupling,
leads to increased fluctuations in $\hat P_m$ (backaction heating) [cf. last term in Eq.~\eqref{VarPcond} and Fig.~\ref{f:Sketch} ({\bf c})].
On the other hand, when the measurement is recorded ($\eta>0$), information about the mechanical state is indirectly acquired, which reduces the uncertainty (variance) as shown by Eq.~\eqref{VarQcond}, eventually resulting in mechanical squeezing $\sigma^2_{X_m}<\tfrac{1}{2}$. 

We show the degree of mechanical squeezing [expressed in $-10\log_{10}(2\sigma^2_{X_m})$ Decibel (dB)] in Fig.~\ref{f:PlotSqRWA}, as a function of the sideband parameter $\kappa/\omega_m$. An optimal value of squeezing emerges for intermediate $\kappa/\omega_m$ as a result of the competition between increasing the  transfer of mechanical information to the light field and  increasing the number of measured photons, which would favour respectively a slower and a faster cavity.   
For a fast cavity $\kappa\gg\omega_m$, we retrieve the adiabatic result $\sigma^2_{X_m}\approx \sigma^2_{X_m,\mathrm{ad}}=\frac{\sqrt{1+4\eta\mathcal{C}(1+2\bar n)}-1}{4\eta\mathcal{C}}$ where we introduced the cooperativity $\mathcal{C}=4g^2/\kappa \gamma$. 
This expression can also be obtained by adiabatically eliminating the cavity mode and considering the resulting effective measurement of $\hat X_m$
\footnote{In the fast-cavity limit the homodyne current can be approximated as \unexpanded{$I(t)dt\approx 2\sqrt{\gamma\eta\mathcal{C}}\langle \hat X_m\rangle dt +dW_t$}
which shows that the problem effectively reduces to the continuous measurement of the mechanical amplitude quadrature}; 
this is the  standard approach for describing the conditional evolution of weakly monitored systems~\cite{Wiseman93, Doherty99,JacobsSteckRev, CMJ2008, TwoModeBAE, Genoni15,QBA18}. 
In the adiabatic regime, decreasing $\kappa$ leads to a larger cooperativity (and hence to a larger effective measurement rate~\cite{CMJ2008})
and determines a steady increase of squeezing.
However, as our solution shows, when this rate becomes smaller than rate at which mechanical information is imprinted onto the light field ($g$), this description becomes inaccurate.
For example, for $g=\kappa=10^{-2}\omega_m$, $\sigma^2_{X_m,\mathrm{ad}}$ overestimates the actually amount of squeezing by approximately  a factor of two (cf.  Fig.~\ref{f:PlotSqRWA}). 
For a slow cavity $\kappa\ll\omega_m$, on the other hand, 
increasing $\kappa$ increases the measurement rate (more photons reaching the detector), which in turn reduces the variance $\sigma^2_{X_m}$.
We can express Eq.~\eqref{VarQcond} in terms of $\mathcal{C}$ and keep only the leading term in the expansion $\mathcal{C}\gg1$, which yields $\sigma^2_{X_m,\mathrm{slow}}=\tfrac{(1+2\bar n)^{3/4}}{(\mathcal{C}\eta)^{1/4}}\sqrt{\gamma/\kappa}$ shown as dashed lines in Fig.~\ref{f:PlotSqRWA}. Once again, this description
loses accuracy when the  cavity loss becomes comparable to the coherent term.
Our exact solution \eqref{VarQcond} interpolates between these two limits and describes a trade-off between two different measurement regimes. 

A more accurate condition for optimal squeezing is obtained from the intersection of the two straight lines in Fig.~\ref{f:PlotSqRWA}
\begin{equation}
\kappa_{\mathrm{opt}}= 4g^{2/3}[\eta\gamma(1+2\bar n)]^{1/3}\,.
\end{equation}
This gives the optimal value of the sideband parameter, which both depends on the rate at which information is transferred to the cavity and on the  thermal decoherence rate.

\emph{Effects of counter-rotating terms.}---We now explore the effect of the counter-rotating (CR) terms appearing in Eq.~\eqref{Hint}. 
As the drift matrix is explicitly time-dependent, we numerically integrate the equations of motion \eqref{Cond2} and consider the long-time limit, when the system settles in a time-periodic steady state. 
The inclusion of the CR terms enables measurement backaction to reach $\hat X_m$ [see Fig.~\ref{f:Sketch} ({\bf d})] and therefore perturb the ideal QND regime.
The consequent reduction of mechanical squeezing can be seen in Fig.~\ref{f:PlotSq} ({\bf a}).
However, such a reduction is accompanied by the emergence of two novel features: (i) the stabilization of optomechanical entanglement to considerably larger values [panels ({\bf b}), ({\bf d})] and (ii) the appearance of squeezing in the cavity quadrature $\hat X_c$ [panel ({\bf c})]. In particular, the presence of CR terms can have a dramatic effect on entanglement, which survives in the steady state, as opposed to the typical entanglement `sudden death' predicted by RWA~\cite{SuddenDeath}. Furthermore, the RWA solution entirely misses intra-cavity squeezing~\cite{SM}.  We thus see that corrections to RWA can lead to qualitatively different features, which is a second major result of our work.

In contrast to unconditional BAE measurements, where CR terms are always detrimental to quantum correlations~\cite{CMJ2008, DanielBAE}, we find that under continuous monitoring quantum correlations can be stronger in their presence. 
Physically, this fact can be traced back to the additional channels opened by CR terms [see Fig.~\ref{f:Sketch} ({\bf d})]. Indeed, as the backaction spreads more, so do the conditioning effects. The inclusion of CR terms favours a \emph{correlated reduction of the uncertainty}, which qualitatively accounts for the emergence of entanglement.
Remarkably, in the strong-coupling regime we observe the joint presence of conditional optical squeezing, mechanical squeezing, and entanglement.
This unusual set of properties has been predicted for the ground state of a pair of bosonic modes in the ultra-strong coupling regime~\cite{Ciuti2005, Ciuti2006} and observed in analog quantum simulation of that model~\cite{UltraStrong}. Continuous monitoring could make the same phenomenology accessible without such stringent experimental requirements.

\emph{Conditional entanglement in a three-mode optomechanical system.}---We now consider two mechanical resonators of frequency $\omega_{m,1}$ and $\omega_{m,2}$ coupled to a common cavity mode, as sketched in Fig.~\ref{f:Sketch} ({\bf b}).
Measuring the output cavity field can induce conditional EPR-like entanglement between them \cite{HammererEPR, TwoModeBAE, HammererAnnalen}.  
Following Ref.~\cite{TwoModeBAE}, we introduce the mean and the relative mechanical frequency
$\omega=(\omega_{m,1}+\omega_{m,2})/2,\, \Omega=(\omega_{m,1}-\omega_{m,2})/2$ (we assume $\omega_{m,1}>\omega_{m,2}$) 
and the collective EPR mechanical variables
\begin{equation}
\hat X_{\pm}=(\hat X_{m,1}\pm\hat X_{m,2})/\sqrt2\, ,\; \hat P_{\pm}=(\hat P_{m,1}\pm\hat P_{m,2})/\sqrt2\,,
\end{equation} 
that satisfy $[\hat X_{\pm},\hat P_{\pm}]=i$, $[\hat X_{\pm},\hat P_{\mp}]=0$. In terms of $\hat X_+$ and $\hat P_-$, all-mechanical entanglement is certified by the violation of Duan's inequality $\sigma_{X_+}^2+\sigma_{P_-}^2\ge1$~\cite{Duan2000}. Amplitude modulation of a resonant drive at 
$\omega$ results in the Hamiltonian
\begin{align}\label{TwoToneQND}
\hat H_I(t)&=\Omega(\hat X_+\hat X_-+\hat P_+\hat P_-)-\sqrt2  g\hat X_c\hat X_+ + \hat H_{\mathrm{CR}}\,.
\end{align}

In the limit $\omega\gg\kappa$, CR terms can be dropped and Eq.~\eqref{TwoToneQND} becomes a perfect two-mode QND interaction~\cite{TwoModeBAE,TsangQND}.
This is due to the fact that $\hat H_{\mathrm{QND}}=\hat H_I(t)-\hat H_{\mathrm{CR}}(t)$ couples $\hat X_+$ and $\hat P_-$ in the same way as for simple harmonic motion, so that the interaction with the cavity turns into a joint continuous measurement of both $\hat X_+$ and $\hat P_-$. Since $\hat X_+$ and $\hat P_-$ commute, they can be simultaneously squeezed by the measurement, while the backaction is confined to 
$\hat P_+$ and $\hat X_-$~\cite{TsangQND}. If their combined uncertainties are reduced below twice the zero-point level, the measurement induces conditional mechanical entanglement, in the form of two-mode squeezing.
 
\begin{figure}[t]
	\centering
\includegraphics[width=0.9\linewidth]{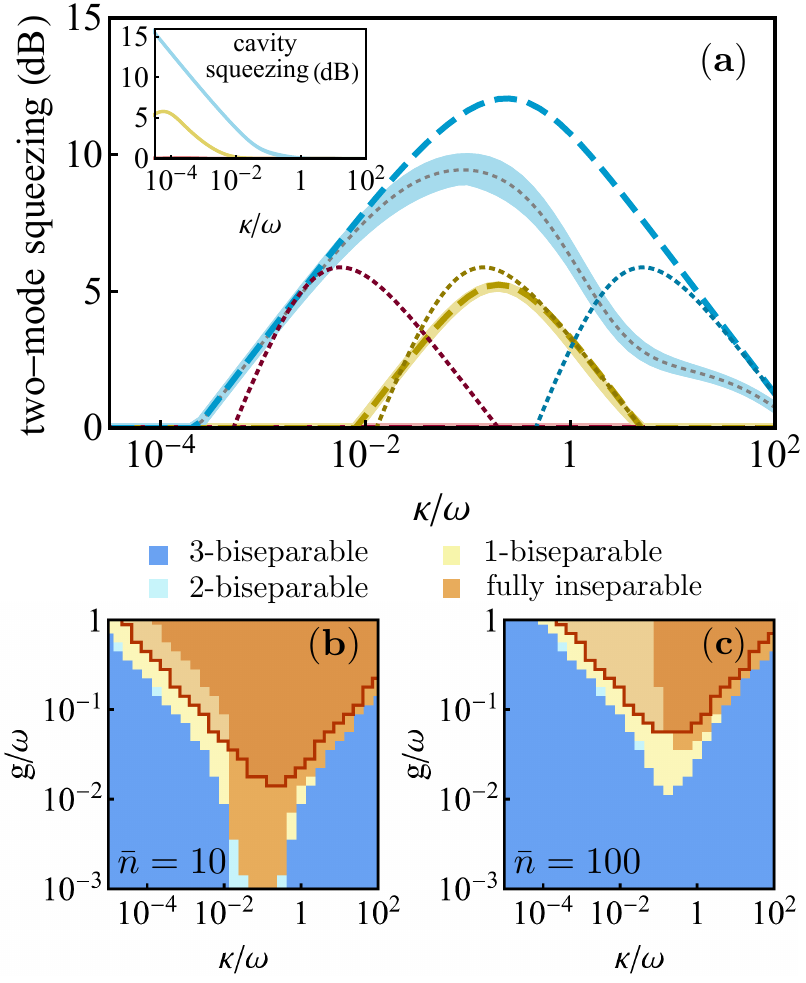}
\caption{
({\bf a}) Mechanical two-mode squeezing (in dB) assuming the RWA (dashed curves), beyond the RWA (lighter  shaded areas) and 
in the adiabatic limit (dotted darker curves). The curves are for $g=0.01\omega$ (red, which is zero), $g=0.05\omega$ (yellow) and $g=0.3\omega$ (cyan). As in Fig.~\ref{f:PlotSq} the dotted gray curve shows the average two-mode squeezing (taken over $2\pi/\omega$). In the inset the conditional cavity squeezing is shown. Other parameters are $\Omega=0.1\omega$, $\gamma=10^{-4}\omega,\, \bar n=10,\,\eta=1$. ({\bf b}) Inseparability structure of the conditional three-mode optomechanical system. 
The shaded region marks the presence of mechanical two-mode squeezing. Other parameters as in ({\bf a}). ({\bf c}) Same as ({\bf b}) except for $\bar n=100$.  
\label{f:PlotTMSq}}
\end{figure}

In Fig.~\ref{f:PlotTMSq} ({\bf a}) we quantify  two-mode squeezing through the violation of Duan's bound. 
We observe a trade-off which can be physically understood as in the single-mode case [cf. Fig.~\ref{f:PlotSqRWA}], although a simple analytic expression [like Eq.~\eqref{VarQcond}] is no longer available.
The effects due to CR terms in Eq.~\eqref{TwoToneQND}, responsible for the reduction of the entanglement and the appearance of cavity squeezing for $g>\kappa$, are akin to our findings for the single-mode case [cf. Fig.~\ref{f:PlotSq} ({\bf a}), ({\bf c})].
We compare our result with the prediction derived in the adiabatic limit (dotted curves, see Ref.~\cite{TwoModeBAE} for the expressions), which is only accurate for $\gamma\ll\Omega, g\ll\kappa\ll\omega$. 
Decreasing the coupling, the adiabatic approximation predicts a constant amount of entanglement, only shifted towards smaller sideband parameters. This prediction can fail dramatically (see red curve), while our theory  correctly quantifies mechanical entanglement in the experimentally relevant good-cavity limit.  
\par
Finally, we study the full conditional dynamics of the three-mode optomechanical system, described by Eq.~\eqref{Cond2}, with the appropriate expressions given in the SM~\cite{SM}. We can determine the separable/entangled nature of the system with respect to all the possible bipartitions, i.e.~$(\hat a\vert\hat b_1\hat b_2)$, $(\hat b_1\vert \hat a \hat b_2)$ and $(\hat b_2\vert \hat a \hat b_1)$, leading to the notion of $k$-biseparable states \cite{ThreeModeSep}.  
In particular, there are states that are entangled for any bipartition of the modes~\footnote{On the other hand, we recall that 3-biseparable states can be either separable or bound entangled states}; these states are called fully inseparable and possess genuine tripartite entanglement. In Fig.~\ref{f:PlotTMSq} ({\bf b}), ({\bf c}) we show the inseparability structure induced by the two-mode QND measurement. We find ample regions where genuinely tripartite entanglement and mechanical two-mode squeezing (marked by the shaded area) coexist, which survive even for large thermal occupation.
Tripartite entanglement in optomechanical devices has been considered in Refs.~\cite{ThreeEntOpto1, ThreeEntOpto2}, however not under continuous monitoring
Most remarkably, our study shows that continuous monitoring can induce non-classical features at every `layer' of the three-mode system: at the single-mode level, the cavity field is squeezed [cf.~inset panel ({\bf a})]; the two-mode mechanical  state is entangled and the optomechanical system as a whole displays genuine multipartite entanglement [Fig.~\ref{f:PlotTMSq} ({\bf b}), ({\bf c})].

\emph{Discussions and conclusions.}---Our results are of direct relevance for ultra-sensitive measurements and feedback-assisted control~\cite{Wilson2015,Rossi2018}. 
Ultra-low dissipation optomechanical sensors featuring large cooperativities operate beyond the adiabatic limit. As we showed, in this regime the enhanced precision achievable (in terms of squeezing) $\sigma^2_{X_m,\mathrm{slow}}\approx \left(\tfrac{8\Gamma_\mathrm{th}^3}{\kappa^2\Gamma_\mathrm{meas}}\right)^{1/4}$ can be related to the thermal decoherence rate $\Gamma_\mathrm{th}=\gamma \bar n$ and to the BAE measurement rate $\Gamma_\mathrm{meas}=4\eta g^2/\kappa$. This provides a benchmark for the performance of linear position sensors working beyond the SQL and a basic requirement for implementing real-time (Markovian) quantum feedback control. 
Our results also extend measurement-based control to multiple degrees of freedom. 
In particular, the possibility of jointly addressing mechanical squeezing, intra-cavity squeezing, and optomechanical entanglement would be useful for ultra-sensitive measurements, e.g. gravitational wave detection~\cite{Ma17}, quantum information processing~\cite{Oussama15} as well as fundamental study of quantum decoherence~\cite{Nimmrichter14,Sam17}. 
  
\emph{Acknowledgments.}---We thank A. Schliesser for discussions at an early stage of the project.
M.~B.~thanks F.~Albarelli, M.~Genoni, and A.~Serafini for useful discussions. D.~M.~acknowledges support by the Horizon 2020 ERC Advanced Grant QUENOCOBA (grant agreement 742102).
A.~N.~acknowledges a University Research Fellowship from the Royal Society and additional support from the Winton Programme for the Physics of Sustainability.
This work was supported by the European UnionÕs Horizon 2020 research and innovation programme under grant agreement No 732894 (FET Proactive HOT).

\bibliography{MyRefs}

\begin{thebibliography}{55}%
\makeatletter
\providecommand \@ifxundefined [1]{%
 \@ifx{#1\undefined}
}%
\providecommand \@ifnum [1]{%
 \ifnum #1\expandafter \@firstoftwo
 \else \expandafter \@secondoftwo
 \fi
}%
\providecommand \@ifx [1]{%
 \ifx #1\expandafter \@firstoftwo
 \else \expandafter \@secondoftwo
 \fi
}%
\providecommand \natexlab [1]{#1}%
\providecommand \enquote  [1]{``#1''}%
\providecommand \bibnamefont  [1]{#1}%
\providecommand \bibfnamefont [1]{#1}%
\providecommand \citenamefont [1]{#1}%
\providecommand \href@noop [0]{\@secondoftwo}%
\providecommand \href [0]{\begingroup \@sanitize@url \@href}%
\providecommand \@href[1]{\@@startlink{#1}\@@href}%
\providecommand \@@href[1]{\endgroup#1\@@endlink}%
\providecommand \@sanitize@url [0]{\catcode `\\12\catcode `\$12\catcode
  `\&12\catcode `\#12\catcode `\^12\catcode `\_12\catcode `\%12\relax}%
\providecommand \@@startlink[1]{}%
\providecommand \@@endlink[0]{}%
\providecommand \url  [0]{\begingroup\@sanitize@url \@url }%
\providecommand \@url [1]{\endgroup\@href {#1}{\urlprefix }}%
\providecommand \urlprefix  [0]{URL }%
\providecommand \Eprint [0]{\href }%
\providecommand \doibase [0]{http://dx.doi.org/}%
\providecommand \selectlanguage [0]{\@gobble}%
\providecommand \bibinfo  [0]{\@secondoftwo}%
\providecommand \bibfield  [0]{\@secondoftwo}%
\providecommand \translation [1]{[#1]}%
\providecommand \BibitemOpen [0]{}%
\providecommand \bibitemStop [0]{}%
\providecommand \bibitemNoStop [0]{.\EOS\space}%
\providecommand \EOS [0]{\spacefactor3000\relax}%
\providecommand \BibitemShut  [1]{\csname bibitem#1\endcsname}%
\let\auto@bib@innerbib\@empty
\bibitem [{\citenamefont {Clerk}\ \emph {et~al.}(2010)\citenamefont {Clerk},
  \citenamefont {Devoret}, \citenamefont {Girvin}, \citenamefont {Marquardt},\
  and\ \citenamefont {Schoelkopf}}]{QuantumNoiseRev}%
  \BibitemOpen
  \bibfield  {author} {\bibinfo {author} {\bibfnamefont {A.~A.}\ \bibnamefont
  {Clerk}}, \bibinfo {author} {\bibfnamefont {M.~H.}\ \bibnamefont {Devoret}},
  \bibinfo {author} {\bibfnamefont {S.~M.}\ \bibnamefont {Girvin}}, \bibinfo
  {author} {\bibfnamefont {F.}~\bibnamefont {Marquardt}}, \ and\ \bibinfo
  {author} {\bibfnamefont {R.~J.}\ \bibnamefont {Schoelkopf}},\ }\href
  {\doibase 10.1103/RevModPhys.82.1155} {\bibfield  {journal} {\bibinfo
  {journal} {Rev. Mod. Phys.}\ }\textbf {\bibinfo {volume} {82}},\ \bibinfo
  {pages} {1155} (\bibinfo {year} {2010})}\BibitemShut {NoStop}%
\bibitem [{\citenamefont {Bocko}\ and\ \citenamefont
  {Onofrio}(1996)}]{OnofrioRev}%
  \BibitemOpen
  \bibfield  {author} {\bibinfo {author} {\bibfnamefont {M.~F.}\ \bibnamefont
  {Bocko}}\ and\ \bibinfo {author} {\bibfnamefont {R.}~\bibnamefont
  {Onofrio}},\ }\href {\doibase 10.1103/RevModPhys.68.755} {\bibfield
  {journal} {\bibinfo  {journal} {Rev. Mod. Phys.}\ }\textbf {\bibinfo {volume}
  {68}},\ \bibinfo {pages} {755} (\bibinfo {year} {1996})}\BibitemShut
  {NoStop}%
\bibitem [{\citenamefont {Braginsky}\ \emph {et~al.}(1980)\citenamefont
  {Braginsky}, \citenamefont {Vorontsov},\ and\ \citenamefont
  {Thorne}}]{Braginsky1}%
  \BibitemOpen
  \bibfield  {author} {\bibinfo {author} {\bibfnamefont {V.~B.}\ \bibnamefont
  {Braginsky}}, \bibinfo {author} {\bibfnamefont {Y.~I.}\ \bibnamefont
  {Vorontsov}}, \ and\ \bibinfo {author} {\bibfnamefont {K.~S.}\ \bibnamefont
  {Thorne}},\ }\href {\doibase 10.1126/science.209.4456.547} {\bibfield
  {journal} {\bibinfo  {journal} {Science}\ }\textbf {\bibinfo {volume}
  {209}},\ \bibinfo {pages} {547} (\bibinfo {year} {1980})}\BibitemShut
  {NoStop}%
\bibitem [{\citenamefont {Thorne}\ \emph {et~al.}(1978)\citenamefont {Thorne},
  \citenamefont {Drever}, \citenamefont {Caves}, \citenamefont {Zimmermann},\
  and\ \citenamefont {Sandberg}}]{QNDThorne78}%
  \BibitemOpen
  \bibfield  {author} {\bibinfo {author} {\bibfnamefont {K.~S.}\ \bibnamefont
  {Thorne}}, \bibinfo {author} {\bibfnamefont {R.~W.~P.}\ \bibnamefont
  {Drever}}, \bibinfo {author} {\bibfnamefont {C.~M.}\ \bibnamefont {Caves}},
  \bibinfo {author} {\bibfnamefont {M.}~\bibnamefont {Zimmermann}}, \ and\
  \bibinfo {author} {\bibfnamefont {V.~D.}\ \bibnamefont {Sandberg}},\ }\href
  {\doibase 10.1103/PhysRevLett.40.667} {\bibfield  {journal} {\bibinfo
  {journal} {Phys. Rev. Lett.}\ }\textbf {\bibinfo {volume} {40}},\ \bibinfo
  {pages} {667} (\bibinfo {year} {1978})}\BibitemShut {NoStop}%
\bibitem [{\citenamefont {Caves}\ \emph {et~al.}(1980)\citenamefont {Caves},
  \citenamefont {Thorne}, \citenamefont {Drever}, \citenamefont {Sandberg},\
  and\ \citenamefont {Zimmermann}}]{WeakMeasRev}%
  \BibitemOpen
  \bibfield  {author} {\bibinfo {author} {\bibfnamefont {C.~M.}\ \bibnamefont
  {Caves}}, \bibinfo {author} {\bibfnamefont {K.~S.}\ \bibnamefont {Thorne}},
  \bibinfo {author} {\bibfnamefont {R.~W.~P.}\ \bibnamefont {Drever}}, \bibinfo
  {author} {\bibfnamefont {V.~D.}\ \bibnamefont {Sandberg}}, \ and\ \bibinfo
  {author} {\bibfnamefont {M.}~\bibnamefont {Zimmermann}},\ }\href {\doibase
  10.1103/RevModPhys.52.341} {\bibfield  {journal} {\bibinfo  {journal} {Rev.
  Mod. Phys.}\ }\textbf {\bibinfo {volume} {52}},\ \bibinfo {pages} {341}
  (\bibinfo {year} {1980})}\BibitemShut {NoStop}%
\bibitem [{\citenamefont {Clerk}\ \emph {et~al.}(2008)\citenamefont {Clerk},
  \citenamefont {Marquardt},\ and\ \citenamefont {Jacobs}}]{CMJ2008}%
  \BibitemOpen
  \bibfield  {author} {\bibinfo {author} {\bibfnamefont {A.~A.}\ \bibnamefont
  {Clerk}}, \bibinfo {author} {\bibfnamefont {F.}~\bibnamefont {Marquardt}}, \
  and\ \bibinfo {author} {\bibfnamefont {K.}~\bibnamefont {Jacobs}},\ }\href
  {http://stacks.iop.org/1367-2630/10/i=9/a=095010} {\bibfield  {journal}
  {\bibinfo  {journal} {New Journal of Physics}\ }\textbf {\bibinfo {volume}
  {10}},\ \bibinfo {pages} {095010} (\bibinfo {year} {2008})}\BibitemShut
  {NoStop}%
\bibitem [{\citenamefont {Hertzberg}\ \emph {et~al.}(2009)\citenamefont
  {Hertzberg}, \citenamefont {Rocheleau}, \citenamefont {Ndukum}, \citenamefont
  {Savva}, \citenamefont {Clerk},\ and\ \citenamefont {Schwab}}]{BAEOpto09}%
  \BibitemOpen
  \bibfield  {author} {\bibinfo {author} {\bibfnamefont {J.~B.}\ \bibnamefont
  {Hertzberg}}, \bibinfo {author} {\bibfnamefont {T.}~\bibnamefont
  {Rocheleau}}, \bibinfo {author} {\bibfnamefont {T.}~\bibnamefont {Ndukum}},
  \bibinfo {author} {\bibfnamefont {M.}~\bibnamefont {Savva}}, \bibinfo
  {author} {\bibfnamefont {A.~A.}\ \bibnamefont {Clerk}}, \ and\ \bibinfo
  {author} {\bibfnamefont {K.~C.}\ \bibnamefont {Schwab}},\ }\href
  {https://doi.org/10.1038/nphys1479} {\bibfield  {journal} {\bibinfo
  {journal} {Nature Physics}\ }\textbf {\bibinfo {volume} {6}},\ \bibinfo
  {pages} {213 EP } (\bibinfo {year} {2009})}\BibitemShut {NoStop}%
\bibitem [{\citenamefont {Suh}\ \emph {et~al.}(2014)\citenamefont {Suh},
  \citenamefont {Weinstein}, \citenamefont {Lei}, \citenamefont {Wollman},
  \citenamefont {Steinke}, \citenamefont {Meystre}, \citenamefont {Clerk},\
  and\ \citenamefont {Schwab}}]{BAEOpto14}%
  \BibitemOpen
  \bibfield  {author} {\bibinfo {author} {\bibfnamefont {J.}~\bibnamefont
  {Suh}}, \bibinfo {author} {\bibfnamefont {A.~J.}\ \bibnamefont {Weinstein}},
  \bibinfo {author} {\bibfnamefont {C.~U.}\ \bibnamefont {Lei}}, \bibinfo
  {author} {\bibfnamefont {E.~E.}\ \bibnamefont {Wollman}}, \bibinfo {author}
  {\bibfnamefont {S.~K.}\ \bibnamefont {Steinke}}, \bibinfo {author}
  {\bibfnamefont {P.}~\bibnamefont {Meystre}}, \bibinfo {author} {\bibfnamefont
  {A.~A.}\ \bibnamefont {Clerk}}, \ and\ \bibinfo {author} {\bibfnamefont
  {K.~C.}\ \bibnamefont {Schwab}},\ }\href@noop {} {\bibfield  {journal}
  {\bibinfo  {journal} {Science}\ }\textbf {\bibinfo {volume} {344}},\ \bibinfo
  {pages} {1262} (\bibinfo {year} {2014})}\BibitemShut {NoStop}%
\bibitem [{\citenamefont {Shomroni}\ \emph {et~al.}(2018)\citenamefont
  {Shomroni}, \citenamefont {Qiu}, \citenamefont {Malz}, \citenamefont
  {Nunnenkamp},\ and\ \citenamefont {Kippenberg}}]{DanielExp18}%
  \BibitemOpen
  \bibfield  {author} {\bibinfo {author} {\bibfnamefont {I.}~\bibnamefont
  {Shomroni}}, \bibinfo {author} {\bibfnamefont {L.}~\bibnamefont {Qiu}},
  \bibinfo {author} {\bibfnamefont {D.}~\bibnamefont {Malz}}, \bibinfo {author}
  {\bibfnamefont {A.}~\bibnamefont {Nunnenkamp}}, \ and\ \bibinfo {author}
  {\bibfnamefont {T.~J.}\ \bibnamefont {Kippenberg}},\ }\href@noop {}
  {\bibfield  {journal} {\bibinfo  {journal} {arXiv:1809.01007}\ } (\bibinfo
  {year} {2018})}\BibitemShut {NoStop}%
\bibitem [{\citenamefont {Vasilakis}\ \emph {et~al.}(2015)\citenamefont
  {Vasilakis}, \citenamefont {Shen}, \citenamefont {Jensen}, \citenamefont
  {Balabas}, \citenamefont {Salart}, \citenamefont {Chen},\ and\ \citenamefont
  {Polzik}}]{BAESpin15}%
  \BibitemOpen
  \bibfield  {author} {\bibinfo {author} {\bibfnamefont {G.}~\bibnamefont
  {Vasilakis}}, \bibinfo {author} {\bibfnamefont {H.}~\bibnamefont {Shen}},
  \bibinfo {author} {\bibfnamefont {K.}~\bibnamefont {Jensen}}, \bibinfo
  {author} {\bibfnamefont {M.}~\bibnamefont {Balabas}}, \bibinfo {author}
  {\bibfnamefont {D.}~\bibnamefont {Salart}}, \bibinfo {author} {\bibfnamefont
  {B.}~\bibnamefont {Chen}}, \ and\ \bibinfo {author} {\bibfnamefont {E.~S.}\
  \bibnamefont {Polzik}},\ }\href@noop {} {\bibfield  {journal} {\bibinfo
  {journal} {Nature Physics}\ }\textbf {\bibinfo {volume} {11}},\ \bibinfo
  {pages} {389 EP } (\bibinfo {year} {2015})}\BibitemShut {NoStop}%
\bibitem [{\citenamefont {Tsang}\ and\ \citenamefont
  {Caves}(2010)}]{TsangCohCanc}%
  \BibitemOpen
  \bibfield  {author} {\bibinfo {author} {\bibfnamefont {M.}~\bibnamefont
  {Tsang}}\ and\ \bibinfo {author} {\bibfnamefont {C.~M.}\ \bibnamefont
  {Caves}},\ }\href {\doibase 10.1103/PhysRevLett.105.123601} {\bibfield
  {journal} {\bibinfo  {journal} {Phys. Rev. Lett.}\ }\textbf {\bibinfo
  {volume} {105}},\ \bibinfo {pages} {123601} (\bibinfo {year}
  {2010})}\BibitemShut {NoStop}%
\bibitem [{\citenamefont {Tsang}\ and\ \citenamefont {Caves}(2012)}]{TsangQND}%
  \BibitemOpen
  \bibfield  {author} {\bibinfo {author} {\bibfnamefont {M.}~\bibnamefont
  {Tsang}}\ and\ \bibinfo {author} {\bibfnamefont {C.~M.}\ \bibnamefont
  {Caves}},\ }\href {\doibase 10.1103/PhysRevX.2.031016} {\bibfield  {journal}
  {\bibinfo  {journal} {Phys. Rev. X}\ }\textbf {\bibinfo {volume} {2}},\
  \bibinfo {pages} {031016} (\bibinfo {year} {2012})}\BibitemShut {NoStop}%
\bibitem [{\citenamefont {Woolley}\ and\ \citenamefont
  {Clerk}(2013)}]{TwoModeBAE}%
  \BibitemOpen
  \bibfield  {author} {\bibinfo {author} {\bibfnamefont {M.~J.}\ \bibnamefont
  {Woolley}}\ and\ \bibinfo {author} {\bibfnamefont {A.~A.}\ \bibnamefont
  {Clerk}},\ }\href {\doibase 10.1103/PhysRevA.87.063846} {\bibfield  {journal}
  {\bibinfo  {journal} {Phys. Rev. A}\ }\textbf {\bibinfo {volume} {87}},\
  \bibinfo {pages} {063846} (\bibinfo {year} {2013})}\BibitemShut {NoStop}%
\bibitem [{\citenamefont {Ockeloen-Korppi}\ \emph {et~al.}(2016)\citenamefont
  {Ockeloen-Korppi}, \citenamefont {Damsk\"agg}, \citenamefont {Pirkkalainen},
  \citenamefont {Clerk}, \citenamefont {Woolley},\ and\ \citenamefont
  {Sillanp\"a\"a}}]{TwoModeBAEExp}%
  \BibitemOpen
  \bibfield  {author} {\bibinfo {author} {\bibfnamefont {C.~F.}\ \bibnamefont
  {Ockeloen-Korppi}}, \bibinfo {author} {\bibfnamefont {E.}~\bibnamefont
  {Damsk\"agg}}, \bibinfo {author} {\bibfnamefont {J.-M.}\ \bibnamefont
  {Pirkkalainen}}, \bibinfo {author} {\bibfnamefont {A.~A.}\ \bibnamefont
  {Clerk}}, \bibinfo {author} {\bibfnamefont {M.~J.}\ \bibnamefont {Woolley}},
  \ and\ \bibinfo {author} {\bibfnamefont {M.~A.}\ \bibnamefont
  {Sillanp\"a\"a}},\ }\href {\doibase 10.1103/PhysRevLett.117.140401}
  {\bibfield  {journal} {\bibinfo  {journal} {Phys. Rev. Lett.}\ }\textbf
  {\bibinfo {volume} {117}},\ \bibinfo {pages} {140401} (\bibinfo {year}
  {2016})}\BibitemShut {NoStop}%
\bibitem [{\citenamefont {Hammerer}\ \emph {et~al.}(2009)\citenamefont
  {Hammerer}, \citenamefont {Aspelmeyer}, \citenamefont {Polzik},\ and\
  \citenamefont {Zoller}}]{HammererEPR}%
  \BibitemOpen
  \bibfield  {author} {\bibinfo {author} {\bibfnamefont {K.}~\bibnamefont
  {Hammerer}}, \bibinfo {author} {\bibfnamefont {M.}~\bibnamefont
  {Aspelmeyer}}, \bibinfo {author} {\bibfnamefont {E.~S.}\ \bibnamefont
  {Polzik}}, \ and\ \bibinfo {author} {\bibfnamefont {P.}~\bibnamefont
  {Zoller}},\ }\href {\doibase 10.1103/PhysRevLett.102.020501} {\bibfield
  {journal} {\bibinfo  {journal} {Phys. Rev. Lett.}\ }\textbf {\bibinfo
  {volume} {102}},\ \bibinfo {pages} {020501} (\bibinfo {year}
  {2009})}\BibitemShut {NoStop}%
\bibitem [{\citenamefont {Zhang}\ \emph {et~al.}(2013)\citenamefont {Zhang},
  \citenamefont {Meystre},\ and\ \citenamefont {Zhang}}]{BAEBEC}%
  \BibitemOpen
  \bibfield  {author} {\bibinfo {author} {\bibfnamefont {K.}~\bibnamefont
  {Zhang}}, \bibinfo {author} {\bibfnamefont {P.}~\bibnamefont {Meystre}}, \
  and\ \bibinfo {author} {\bibfnamefont {W.}~\bibnamefont {Zhang}},\ }\href
  {\doibase 10.1103/PhysRevA.88.043632} {\bibfield  {journal} {\bibinfo
  {journal} {Phys. Rev. A}\ }\textbf {\bibinfo {volume} {88}},\ \bibinfo
  {pages} {043632} (\bibinfo {year} {2013})}\BibitemShut {NoStop}%
\bibitem [{\citenamefont {M{\o}ller}\ \emph {et~al.}(2017)\citenamefont
  {M{\o}ller}, \citenamefont {Thomas}, \citenamefont {Vasilakis}, \citenamefont
  {Zeuthen}, \citenamefont {Tsaturyan}, \citenamefont {Balabas}, \citenamefont
  {Jensen}, \citenamefont {Schliesser}, \citenamefont {Hammerer},\ and\
  \citenamefont {Polzik}}]{HybridBAE}%
  \BibitemOpen
  \bibfield  {author} {\bibinfo {author} {\bibfnamefont {C.~B.}\ \bibnamefont
  {M{\o}ller}}, \bibinfo {author} {\bibfnamefont {R.~A.}\ \bibnamefont
  {Thomas}}, \bibinfo {author} {\bibfnamefont {G.}~\bibnamefont {Vasilakis}},
  \bibinfo {author} {\bibfnamefont {E.}~\bibnamefont {Zeuthen}}, \bibinfo
  {author} {\bibfnamefont {Y.}~\bibnamefont {Tsaturyan}}, \bibinfo {author}
  {\bibfnamefont {M.}~\bibnamefont {Balabas}}, \bibinfo {author} {\bibfnamefont
  {K.}~\bibnamefont {Jensen}}, \bibinfo {author} {\bibfnamefont
  {A.}~\bibnamefont {Schliesser}}, \bibinfo {author} {\bibfnamefont
  {K.}~\bibnamefont {Hammerer}}, \ and\ \bibinfo {author} {\bibfnamefont
  {E.~S.}\ \bibnamefont {Polzik}},\ }\href
  {https://doi.org/10.1038/nature22980} {\bibfield  {journal} {\bibinfo
  {journal} {Nature}\ }\textbf {\bibinfo {volume} {547}},\ \bibinfo {pages}
  {191 EP } (\bibinfo {year} {2017})}\BibitemShut {NoStop}%
\bibitem [{\citenamefont {Wilson}\ \emph {et~al.}(2015)\citenamefont {Wilson},
  \citenamefont {Sudhir}, \citenamefont {Piro}, \citenamefont {Schilling},
  \citenamefont {Ghadimi},\ and\ \citenamefont {Kippenberg}}]{Wilson2015}%
  \BibitemOpen
  \bibfield  {author} {\bibinfo {author} {\bibfnamefont {D.~J.}\ \bibnamefont
  {Wilson}}, \bibinfo {author} {\bibfnamefont {V.}~\bibnamefont {Sudhir}},
  \bibinfo {author} {\bibfnamefont {N.}~\bibnamefont {Piro}}, \bibinfo {author}
  {\bibfnamefont {R.}~\bibnamefont {Schilling}}, \bibinfo {author}
  {\bibfnamefont {A.}~\bibnamefont {Ghadimi}}, \ and\ \bibinfo {author}
  {\bibfnamefont {T.~J.}\ \bibnamefont {Kippenberg}},\ }\href
  {https://doi.org/10.1038/nature14672} {\bibfield  {journal} {\bibinfo
  {journal} {Nature}\ }\textbf {\bibinfo {volume} {524}},\ \bibinfo {pages}
  {325 EP } (\bibinfo {year} {2015})}\BibitemShut {NoStop}%
\bibitem [{\citenamefont {Sudhir}\ \emph {et~al.}(2017)\citenamefont {Sudhir},
  \citenamefont {Wilson}, \citenamefont {Schilling}, \citenamefont {Sch\"utz},
  \citenamefont {Fedorov}, \citenamefont {Ghadimi}, \citenamefont
  {Nunnenkamp},\ and\ \citenamefont {Kippenberg}}]{Shudir2017}%
  \BibitemOpen
  \bibfield  {author} {\bibinfo {author} {\bibfnamefont {V.}~\bibnamefont
  {Sudhir}}, \bibinfo {author} {\bibfnamefont {D.~J.}\ \bibnamefont {Wilson}},
  \bibinfo {author} {\bibfnamefont {R.}~\bibnamefont {Schilling}}, \bibinfo
  {author} {\bibfnamefont {H.}~\bibnamefont {Sch\"utz}}, \bibinfo {author}
  {\bibfnamefont {S.~A.}\ \bibnamefont {Fedorov}}, \bibinfo {author}
  {\bibfnamefont {A.~H.}\ \bibnamefont {Ghadimi}}, \bibinfo {author}
  {\bibfnamefont {A.}~\bibnamefont {Nunnenkamp}}, \ and\ \bibinfo {author}
  {\bibfnamefont {T.~J.}\ \bibnamefont {Kippenberg}},\ }\href {\doibase
  10.1103/PhysRevX.7.011001} {\bibfield  {journal} {\bibinfo  {journal} {Phys.
  Rev. X}\ }\textbf {\bibinfo {volume} {7}},\ \bibinfo {pages} {011001}
  (\bibinfo {year} {2017})}\BibitemShut {NoStop}%
\bibitem [{\citenamefont {Rossi}\ \emph {et~al.}(2017)\citenamefont {Rossi},
  \citenamefont {Kralj}, \citenamefont {Zippilli}, \citenamefont {Natali},
  \citenamefont {Borrielli}, \citenamefont {Pandraud}, \citenamefont {Serra},
  \citenamefont {Di~Giuseppe},\ and\ \citenamefont {Vitali}}]{Rossi2017}%
  \BibitemOpen
  \bibfield  {author} {\bibinfo {author} {\bibfnamefont {M.}~\bibnamefont
  {Rossi}}, \bibinfo {author} {\bibfnamefont {N.}~\bibnamefont {Kralj}},
  \bibinfo {author} {\bibfnamefont {S.}~\bibnamefont {Zippilli}}, \bibinfo
  {author} {\bibfnamefont {R.}~\bibnamefont {Natali}}, \bibinfo {author}
  {\bibfnamefont {A.}~\bibnamefont {Borrielli}}, \bibinfo {author}
  {\bibfnamefont {G.}~\bibnamefont {Pandraud}}, \bibinfo {author}
  {\bibfnamefont {E.}~\bibnamefont {Serra}}, \bibinfo {author} {\bibfnamefont
  {G.}~\bibnamefont {Di~Giuseppe}}, \ and\ \bibinfo {author} {\bibfnamefont
  {D.}~\bibnamefont {Vitali}},\ }\href {\doibase
  10.1103/PhysRevLett.119.123603} {\bibfield  {journal} {\bibinfo  {journal}
  {Phys. Rev. Lett.}\ }\textbf {\bibinfo {volume} {119}},\ \bibinfo {pages}
  {123603} (\bibinfo {year} {2017})}\BibitemShut {NoStop}%
\bibitem [{\citenamefont {Rossi}\ \emph
  {et~al.}(2018{\natexlab{a}})\citenamefont {Rossi}, \citenamefont {Mason},
  \citenamefont {Chen}, \citenamefont {Tsaturyan},\ and\ \citenamefont
  {Schliesser}}]{Rossi2018}%
  \BibitemOpen
  \bibfield  {author} {\bibinfo {author} {\bibfnamefont {M.}~\bibnamefont
  {Rossi}}, \bibinfo {author} {\bibfnamefont {D.}~\bibnamefont {Mason}},
  \bibinfo {author} {\bibfnamefont {J.}~\bibnamefont {Chen}}, \bibinfo {author}
  {\bibfnamefont {Y.}~\bibnamefont {Tsaturyan}}, \ and\ \bibinfo {author}
  {\bibfnamefont {A.}~\bibnamefont {Schliesser}},\ }\href {\doibase
  10.1038/s41586-018-0643-8} {\bibfield  {journal} {\bibinfo  {journal}
  {Nature}\ }\textbf {\bibinfo {volume} {563}},\ \bibinfo {pages} {53}
  (\bibinfo {year} {2018}{\natexlab{a}})}\BibitemShut {NoStop}%
\bibitem [{\citenamefont {Wiseman}\ and\ \citenamefont
  {Milburn}(1993)}]{Wiseman93}%
  \BibitemOpen
  \bibfield  {author} {\bibinfo {author} {\bibfnamefont {H.~M.}\ \bibnamefont
  {Wiseman}}\ and\ \bibinfo {author} {\bibfnamefont {G.~J.}\ \bibnamefont
  {Milburn}},\ }\href {\doibase 10.1103/PhysRevA.47.642} {\bibfield  {journal}
  {\bibinfo  {journal} {Phys. Rev. A}\ }\textbf {\bibinfo {volume} {47}},\
  \bibinfo {pages} {642} (\bibinfo {year} {1993})}\BibitemShut {NoStop}%
\bibitem [{\citenamefont {Wiseman}\ and\ \citenamefont
  {Milburn}(2009)}]{QuantumMeasBook}%
  \BibitemOpen
  \bibfield  {author} {\bibinfo {author} {\bibfnamefont {H.~M.}\ \bibnamefont
  {Wiseman}}\ and\ \bibinfo {author} {\bibfnamefont {G.~J.}\ \bibnamefont
  {Milburn}},\ }\href {\doibase 10.1017/CBO9780511813948} {\emph {\bibinfo
  {title} {Quantum Measurement and Control}}}\ (\bibinfo  {publisher}
  {Cambridge University Press},\ \bibinfo {year} {2009})\BibitemShut {NoStop}%
\bibitem [{\citenamefont {Jacobs}\ and\ \citenamefont
  {Steck}(2006)}]{JacobsSteckRev}%
  \BibitemOpen
  \bibfield  {author} {\bibinfo {author} {\bibfnamefont {K.}~\bibnamefont
  {Jacobs}}\ and\ \bibinfo {author} {\bibfnamefont {D.~A.}\ \bibnamefont
  {Steck}},\ }\href {\doibase 10.1080/00107510601101934} {\bibfield  {journal}
  {\bibinfo  {journal} {Contemporary Physics}\ }\textbf {\bibinfo {volume}
  {47}},\ \bibinfo {pages} {279} (\bibinfo {year} {2006})}\BibitemShut
  {NoStop}%
\bibitem [{\citenamefont {Kwon}\ \emph {et~al.}(2019)\citenamefont {Kwon},
  \citenamefont {Tan}, \citenamefont {Volkoff},\ and\ \citenamefont
  {Jeong}}]{NonClassMetrology}%
  \BibitemOpen
  \bibfield  {author} {\bibinfo {author} {\bibfnamefont {H.}~\bibnamefont
  {Kwon}}, \bibinfo {author} {\bibfnamefont {K.~C.}\ \bibnamefont {Tan}},
  \bibinfo {author} {\bibfnamefont {T.}~\bibnamefont {Volkoff}}, \ and\
  \bibinfo {author} {\bibfnamefont {H.}~\bibnamefont {Jeong}},\ }\href
  {\doibase 10.1103/PhysRevLett.122.040503} {\bibfield  {journal} {\bibinfo
  {journal} {Phys. Rev. Lett.}\ }\textbf {\bibinfo {volume} {122}},\ \bibinfo
  {pages} {040503} (\bibinfo {year} {2019})}\BibitemShut {NoStop}%
\bibitem [{\citenamefont {Garbe}\ \emph {et~al.}(2019)\citenamefont {Garbe},
  \citenamefont {Felicetti}, \citenamefont {Milman}, \citenamefont {Coudreau},\
  and\ \citenamefont {Keller}}]{SqueezingMetrology}%
  \BibitemOpen
  \bibfield  {author} {\bibinfo {author} {\bibfnamefont {L.}~\bibnamefont
  {Garbe}}, \bibinfo {author} {\bibfnamefont {S.}~\bibnamefont {Felicetti}},
  \bibinfo {author} {\bibfnamefont {P.}~\bibnamefont {Milman}}, \bibinfo
  {author} {\bibfnamefont {T.}~\bibnamefont {Coudreau}}, \ and\ \bibinfo
  {author} {\bibfnamefont {A.}~\bibnamefont {Keller}},\ }\href {\doibase
  10.1103/PhysRevA.99.043815} {\bibfield  {journal} {\bibinfo  {journal} {Phys.
  Rev. A}\ }\textbf {\bibinfo {volume} {99}},\ \bibinfo {pages} {043815}
  (\bibinfo {year} {2019})}\BibitemShut {NoStop}%
\bibitem [{\citenamefont {Weedbrook}\ \emph {et~al.}(2012)\citenamefont
  {Weedbrook}, \citenamefont {Pirandola}, \citenamefont {Garc\'{\i}a-Patr\'on},
  \citenamefont {Cerf}, \citenamefont {Ralph}, \citenamefont {Shapiro},\ and\
  \citenamefont {Lloyd}}]{GaussRev}%
  \BibitemOpen
  \bibfield  {author} {\bibinfo {author} {\bibfnamefont {C.}~\bibnamefont
  {Weedbrook}}, \bibinfo {author} {\bibfnamefont {S.}~\bibnamefont
  {Pirandola}}, \bibinfo {author} {\bibfnamefont {R.}~\bibnamefont
  {Garc\'{\i}a-Patr\'on}}, \bibinfo {author} {\bibfnamefont {N.~J.}\
  \bibnamefont {Cerf}}, \bibinfo {author} {\bibfnamefont {T.~C.}\ \bibnamefont
  {Ralph}}, \bibinfo {author} {\bibfnamefont {J.~H.}\ \bibnamefont {Shapiro}},
  \ and\ \bibinfo {author} {\bibfnamefont {S.}~\bibnamefont {Lloyd}},\ }\href
  {\doibase 10.1103/RevModPhys.84.621} {\bibfield  {journal} {\bibinfo
  {journal} {Rev. Mod. Phys.}\ }\textbf {\bibinfo {volume} {84}},\ \bibinfo
  {pages} {621} (\bibinfo {year} {2012})}\BibitemShut {NoStop}%
\bibitem [{\citenamefont {Houhou}\ \emph {et~al.}(2015)\citenamefont {Houhou},
  \citenamefont {Aissaoui},\ and\ \citenamefont {Ferraro}}]{Oussama15}%
  \BibitemOpen
  \bibfield  {author} {\bibinfo {author} {\bibfnamefont {O.}~\bibnamefont
  {Houhou}}, \bibinfo {author} {\bibfnamefont {H.}~\bibnamefont {Aissaoui}}, \
  and\ \bibinfo {author} {\bibfnamefont {A.}~\bibnamefont {Ferraro}},\ }\href
  {\doibase 10.1103/PhysRevA.92.063843} {\bibfield  {journal} {\bibinfo
  {journal} {Phys. Rev. A}\ }\textbf {\bibinfo {volume} {92}},\ \bibinfo
  {pages} {063843} (\bibinfo {year} {2015})}\BibitemShut {NoStop}%
\bibitem [{\citenamefont {Houhou}\ \emph {et~al.}(2018)\citenamefont {Houhou},
  \citenamefont {Moore}, \citenamefont {Bose},\ and\ \citenamefont
  {Ferraro}}]{Oussama18}%
  \BibitemOpen
  \bibfield  {author} {\bibinfo {author} {\bibfnamefont {O.}~\bibnamefont
  {Houhou}}, \bibinfo {author} {\bibfnamefont {D.~W.}\ \bibnamefont {Moore}},
  \bibinfo {author} {\bibfnamefont {S.}~\bibnamefont {Bose}}, \ and\ \bibinfo
  {author} {\bibfnamefont {A.}~\bibnamefont {Ferraro}},\ }\href@noop {}
  {\bibfield  {journal} {\bibinfo  {journal} {arXiv:1809.09733}\ } (\bibinfo
  {year} {2018})}\BibitemShut {NoStop}%
\bibitem [{\citenamefont {Doherty}\ and\ \citenamefont
  {Jacobs}(1999)}]{Doherty99}%
  \BibitemOpen
  \bibfield  {author} {\bibinfo {author} {\bibfnamefont {A.~C.}\ \bibnamefont
  {Doherty}}\ and\ \bibinfo {author} {\bibfnamefont {K.}~\bibnamefont
  {Jacobs}},\ }\href {\doibase 10.1103/PhysRevA.60.2700} {\bibfield  {journal}
  {\bibinfo  {journal} {Phys. Rev. A}\ }\textbf {\bibinfo {volume} {60}},\
  \bibinfo {pages} {2700} (\bibinfo {year} {1999})}\BibitemShut {NoStop}%
\bibitem [{\citenamefont {Rossi}\ \emph
  {et~al.}(2018{\natexlab{b}})\citenamefont {Rossi}, \citenamefont {Mason},
  \citenamefont {Chen},\ and\ \citenamefont {Schliesser}}]{QBA18}%
  \BibitemOpen
  \bibfield  {author} {\bibinfo {author} {\bibfnamefont {M.}~\bibnamefont
  {Rossi}}, \bibinfo {author} {\bibfnamefont {D.}~\bibnamefont {Mason}},
  \bibinfo {author} {\bibfnamefont {J.}~\bibnamefont {Chen}}, \ and\ \bibinfo
  {author} {\bibfnamefont {A.}~\bibnamefont {Schliesser}},\ }\href@noop {}
  {\bibfield  {journal} {\bibinfo  {journal} {arXiv:1812.00928}\ } (\bibinfo
  {year} {2018}{\natexlab{b}})}\BibitemShut {NoStop}%
\bibitem [{\citenamefont {Hacohen-Gourgy}\ \emph {et~al.}(2016)\citenamefont
  {Hacohen-Gourgy}, \citenamefont {S.~Martin}, \citenamefont {Flurin},
  \citenamefont {Ramasesh}, \citenamefont {Whaley},\ and\ \citenamefont
  {Siddiqi}}]{Hacohen-Gourgy2016}%
  \BibitemOpen
  \bibfield  {author} {\bibinfo {author} {\bibfnamefont {S.}~\bibnamefont
  {Hacohen-Gourgy}}, \bibinfo {author} {\bibfnamefont {L.}~\bibnamefont
  {S.~Martin}}, \bibinfo {author} {\bibfnamefont {E.}~\bibnamefont {Flurin}},
  \bibinfo {author} {\bibfnamefont {V.~V.}\ \bibnamefont {Ramasesh}}, \bibinfo
  {author} {\bibfnamefont {K.~B.}\ \bibnamefont {Whaley}}, \ and\ \bibinfo
  {author} {\bibfnamefont {I.}~\bibnamefont {Siddiqi}},\ }\href {\doibase
  10.1038/nature19762} {\bibfield  {journal} {\bibinfo  {journal} {Nature}\ }
  (\bibinfo {year} {2016}),\ 10.1038/nature19762}\BibitemShut {NoStop}%
\bibitem [{\citenamefont {Chantasri}\ \emph {et~al.}(2016)\citenamefont
  {Chantasri}, \citenamefont {Kimchi-Schwartz}, \citenamefont {Roch},
  \citenamefont {Siddiqi},\ and\ \citenamefont {Jordan}}]{Chantasri2016}%
  \BibitemOpen
  \bibfield  {author} {\bibinfo {author} {\bibfnamefont {A.}~\bibnamefont
  {Chantasri}}, \bibinfo {author} {\bibfnamefont {M.~E.}\ \bibnamefont
  {Kimchi-Schwartz}}, \bibinfo {author} {\bibfnamefont {N.}~\bibnamefont
  {Roch}}, \bibinfo {author} {\bibfnamefont {I.}~\bibnamefont {Siddiqi}}, \
  and\ \bibinfo {author} {\bibfnamefont {A.~N.}\ \bibnamefont {Jordan}},\
  }\href {\doibase 10.1103/PhysRevX.6.041052} {\bibfield  {journal} {\bibinfo
  {journal} {Phys. Rev. X}\ }\textbf {\bibinfo {volume} {6}},\ \bibinfo {pages}
  {041052} (\bibinfo {year} {2016})}\BibitemShut {NoStop}%
\bibitem [{\citenamefont {Murch}\ \emph {et~al.}(2013)\citenamefont {Murch},
  \citenamefont {Weber}, \citenamefont {Macklin},\ and\ \citenamefont
  {Siddiqi}}]{Murch2013}%
  \BibitemOpen
  \bibfield  {author} {\bibinfo {author} {\bibfnamefont {K.~W.}\ \bibnamefont
  {Murch}}, \bibinfo {author} {\bibfnamefont {S.~J.}\ \bibnamefont {Weber}},
  \bibinfo {author} {\bibfnamefont {C.}~\bibnamefont {Macklin}}, \ and\
  \bibinfo {author} {\bibfnamefont {I.}~\bibnamefont {Siddiqi}},\ }\href
  {https://doi.org/10.1038/nature12539} {\bibfield  {journal} {\bibinfo
  {journal} {Nature}\ }\textbf {\bibinfo {volume} {502}},\ \bibinfo {pages}
  {211 EP } (\bibinfo {year} {2013})}\BibitemShut {NoStop}%
\bibitem [{\citenamefont {Aspelmeyer}\ \emph {et~al.}(2014)\citenamefont
  {Aspelmeyer}, \citenamefont {Kippenberg},\ and\ \citenamefont
  {Marquardt}}]{OptoRev}%
  \BibitemOpen
  \bibfield  {author} {\bibinfo {author} {\bibfnamefont {M.}~\bibnamefont
  {Aspelmeyer}}, \bibinfo {author} {\bibfnamefont {T.~J.}\ \bibnamefont
  {Kippenberg}}, \ and\ \bibinfo {author} {\bibfnamefont {F.}~\bibnamefont
  {Marquardt}},\ }\href {\doibase 10.1103/RevModPhys.86.1391} {\bibfield
  {journal} {\bibinfo  {journal} {Rev. Mod. Phys.}\ }\textbf {\bibinfo {volume}
  {86}},\ \bibinfo {pages} {1391} (\bibinfo {year} {2014})}\BibitemShut
  {NoStop}%
\bibitem [{SM()}]{SM}%
  \BibitemOpen
  \href@noop {} {}\bibinfo {note} {See Supplementary Material, which contains
  Ref.~\cite{Mari2009}}\BibitemShut {NoStop}%
\bibitem [{\citenamefont {Genoni}\ \emph {et~al.}(2016)\citenamefont {Genoni},
  \citenamefont {Lami},\ and\ \citenamefont {Serafini}}]{Marco16}%
  \BibitemOpen
  \bibfield  {author} {\bibinfo {author} {\bibfnamefont {M.~G.}\ \bibnamefont
  {Genoni}}, \bibinfo {author} {\bibfnamefont {L.}~\bibnamefont {Lami}}, \ and\
  \bibinfo {author} {\bibfnamefont {A.}~\bibnamefont {Serafini}},\ }\href
  {\doibase 10.1080/00107514.2015.1125624} {\bibfield  {journal} {\bibinfo
  {journal} {Contemporary Physics}\ }\textbf {\bibinfo {volume} {57}},\
  \bibinfo {pages} {331} (\bibinfo {year} {2016})}\BibitemShut {NoStop}%
\bibitem [{\citenamefont {Serafini}(2017)}]{AlessioBook}%
  \BibitemOpen
  \bibfield  {author} {\bibinfo {author} {\bibfnamefont {A.}~\bibnamefont
  {Serafini}},\ }\href {\doibase 10.1201/9781315118727} {\emph {\bibinfo
  {title} {Quantum Continuous Variables: A Primer of Theoretical Methods}}}\
  (\bibinfo  {publisher} {CRC Press},\ \bibinfo {year} {2017})\BibitemShut
  {NoStop}%
\bibitem [{Note1()}]{Note1}%
  \BibitemOpen
  \bibinfo {note} {In the fast-cavity limit the homodyne current can be
  approximated as $I(t)dt\approx 2\sqrt {\gamma \eta \mathcal {C}}\langle \hat
  X_m\rangle dt +dW_t$ which shows that the problem effectively reduces to the
  continuous measurement of the mechanical amplitude quadrature}\BibitemShut
  {NoStop}%
\bibitem [{\citenamefont {Genoni}\ \emph {et~al.}(2015)\citenamefont {Genoni},
  \citenamefont {Zhang}, \citenamefont {Millen}, \citenamefont {Barker},\ and\
  \citenamefont {Serafini}}]{Genoni15}%
  \BibitemOpen
  \bibfield  {author} {\bibinfo {author} {\bibfnamefont {M.~G.}\ \bibnamefont
  {Genoni}}, \bibinfo {author} {\bibfnamefont {J.}~\bibnamefont {Zhang}},
  \bibinfo {author} {\bibfnamefont {J.}~\bibnamefont {Millen}}, \bibinfo
  {author} {\bibfnamefont {P.~F.}\ \bibnamefont {Barker}}, \ and\ \bibinfo
  {author} {\bibfnamefont {A.}~\bibnamefont {Serafini}},\ }\href {\doibase
  10.1088/1367-2630/17/7/073019} {\bibfield  {journal} {\bibinfo  {journal}
  {New Journal of Physics}\ }\textbf {\bibinfo {volume} {17}},\ \bibinfo
  {pages} {073019} (\bibinfo {year} {2015})}\BibitemShut {NoStop}%
\bibitem [{\citenamefont {Yu}\ and\ \citenamefont
  {Eberly}(2009)}]{SuddenDeath}%
  \BibitemOpen
  \bibfield  {author} {\bibinfo {author} {\bibfnamefont {T.}~\bibnamefont
  {Yu}}\ and\ \bibinfo {author} {\bibfnamefont {J.~H.}\ \bibnamefont
  {Eberly}},\ }\href {\doibase 10.1126/science.1167343} {\bibfield  {journal}
  {\bibinfo  {journal} {Science}\ }\textbf {\bibinfo {volume} {323}},\ \bibinfo
  {pages} {598} (\bibinfo {year} {2009})}\BibitemShut {NoStop}%
\bibitem [{\citenamefont {Malz}\ and\ \citenamefont
  {Nunnenkamp}(2016)}]{DanielBAE}%
  \BibitemOpen
  \bibfield  {author} {\bibinfo {author} {\bibfnamefont {D.}~\bibnamefont
  {Malz}}\ and\ \bibinfo {author} {\bibfnamefont {A.}~\bibnamefont
  {Nunnenkamp}},\ }\href {\doibase 10.1103/PhysRevA.94.053820} {\bibfield
  {journal} {\bibinfo  {journal} {Phys. Rev. A}\ }\textbf {\bibinfo {volume}
  {94}},\ \bibinfo {pages} {053820} (\bibinfo {year} {2016})}\BibitemShut
  {NoStop}%
\bibitem [{\citenamefont {Ciuti}\ and\ \citenamefont
  {Carusotto}(2006{\natexlab{a}})}]{Ciuti2005}%
  \BibitemOpen
  \bibfield  {author} {\bibinfo {author} {\bibfnamefont {C.}~\bibnamefont
  {Ciuti}}\ and\ \bibinfo {author} {\bibfnamefont {I.}~\bibnamefont
  {Carusotto}},\ }\href {\doibase 10.1103/PhysRevA.74.033811} {\bibfield
  {journal} {\bibinfo  {journal} {Phys. Rev. A}\ }\textbf {\bibinfo {volume}
  {74}},\ \bibinfo {pages} {033811} (\bibinfo {year}
  {2006}{\natexlab{a}})}\BibitemShut {NoStop}%
\bibitem [{\citenamefont {Ciuti}\ and\ \citenamefont
  {Carusotto}(2006{\natexlab{b}})}]{Ciuti2006}%
  \BibitemOpen
  \bibfield  {author} {\bibinfo {author} {\bibfnamefont {C.}~\bibnamefont
  {Ciuti}}\ and\ \bibinfo {author} {\bibfnamefont {I.}~\bibnamefont
  {Carusotto}},\ }\href {\doibase 10.1103/PhysRevA.74.033811} {\bibfield
  {journal} {\bibinfo  {journal} {Phys. Rev. A}\ }\textbf {\bibinfo {volume}
  {74}},\ \bibinfo {pages} {033811} (\bibinfo {year}
  {2006}{\natexlab{b}})}\BibitemShut {NoStop}%
\bibitem [{\citenamefont {Markovi\ifmmode~\acute{c}\else \'{c}\fi{}}\ \emph
  {et~al.}(2018)\citenamefont {Markovi\ifmmode~\acute{c}\else \'{c}\fi{}},
  \citenamefont {Jezouin}, \citenamefont {Ficheux}, \citenamefont
  {Fedortchenko}, \citenamefont {Felicetti}, \citenamefont {Coudreau},
  \citenamefont {Milman}, \citenamefont {Leghtas},\ and\ \citenamefont
  {Huard}}]{UltraStrong}%
  \BibitemOpen
  \bibfield  {author} {\bibinfo {author} {\bibfnamefont {D.}~\bibnamefont
  {Markovi\ifmmode~\acute{c}\else \'{c}\fi{}}}, \bibinfo {author}
  {\bibfnamefont {S.}~\bibnamefont {Jezouin}}, \bibinfo {author} {\bibfnamefont
  {Q.}~\bibnamefont {Ficheux}}, \bibinfo {author} {\bibfnamefont
  {S.}~\bibnamefont {Fedortchenko}}, \bibinfo {author} {\bibfnamefont
  {S.}~\bibnamefont {Felicetti}}, \bibinfo {author} {\bibfnamefont
  {T.}~\bibnamefont {Coudreau}}, \bibinfo {author} {\bibfnamefont
  {P.}~\bibnamefont {Milman}}, \bibinfo {author} {\bibfnamefont
  {Z.}~\bibnamefont {Leghtas}}, \ and\ \bibinfo {author} {\bibfnamefont
  {B.}~\bibnamefont {Huard}},\ }\href {\doibase 10.1103/PhysRevLett.121.040505}
  {\bibfield  {journal} {\bibinfo  {journal} {Phys. Rev. Lett.}\ }\textbf
  {\bibinfo {volume} {121}},\ \bibinfo {pages} {040505} (\bibinfo {year}
  {2018})}\BibitemShut {NoStop}%
\bibitem [{\citenamefont {Polzik}\ and\ \citenamefont
  {Hammerer}(2015)}]{HammererAnnalen}%
  \BibitemOpen
  \bibfield  {author} {\bibinfo {author} {\bibfnamefont {E.~S.}\ \bibnamefont
  {Polzik}}\ and\ \bibinfo {author} {\bibfnamefont {K.}~\bibnamefont
  {Hammerer}},\ }\href {\doibase 10.1002/andp.201400099} {\bibfield  {journal}
  {\bibinfo  {journal} {Annalen der Physik}\ }\textbf {\bibinfo {volume}
  {527}},\ \bibinfo {pages} {A15} (\bibinfo {year} {2015})}\BibitemShut
  {NoStop}%
\bibitem [{\citenamefont {Duan}\ \emph {et~al.}(2000)\citenamefont {Duan},
  \citenamefont {Giedke}, \citenamefont {Cirac},\ and\ \citenamefont
  {Zoller}}]{Duan2000}%
  \BibitemOpen
  \bibfield  {author} {\bibinfo {author} {\bibfnamefont {L.-M.}\ \bibnamefont
  {Duan}}, \bibinfo {author} {\bibfnamefont {G.}~\bibnamefont {Giedke}},
  \bibinfo {author} {\bibfnamefont {J.~I.}\ \bibnamefont {Cirac}}, \ and\
  \bibinfo {author} {\bibfnamefont {P.}~\bibnamefont {Zoller}},\ }\href
  {\doibase 10.1103/PhysRevLett.84.2722} {\bibfield  {journal} {\bibinfo
  {journal} {Phys. Rev. Lett.}\ }\textbf {\bibinfo {volume} {84}},\ \bibinfo
  {pages} {2722} (\bibinfo {year} {2000})}\BibitemShut {NoStop}%
\bibitem [{\citenamefont {Giedke}\ \emph {et~al.}(2001)\citenamefont {Giedke},
  \citenamefont {Kraus}, \citenamefont {Lewenstein},\ and\ \citenamefont
  {Cirac}}]{ThreeModeSep}%
  \BibitemOpen
  \bibfield  {author} {\bibinfo {author} {\bibfnamefont {G.}~\bibnamefont
  {Giedke}}, \bibinfo {author} {\bibfnamefont {B.}~\bibnamefont {Kraus}},
  \bibinfo {author} {\bibfnamefont {M.}~\bibnamefont {Lewenstein}}, \ and\
  \bibinfo {author} {\bibfnamefont {J.~I.}\ \bibnamefont {Cirac}},\ }\href
  {\doibase 10.1103/PhysRevA.64.052303} {\bibfield  {journal} {\bibinfo
  {journal} {Phys. Rev. A}\ }\textbf {\bibinfo {volume} {64}},\ \bibinfo
  {pages} {052303} (\bibinfo {year} {2001})}\BibitemShut {NoStop}%
\bibitem [{Note2()}]{Note2}%
  \BibitemOpen
  \bibinfo {note} {On the other hand, we recall that 3-biseparable states can
  be either separable or bound entangled states}\BibitemShut {NoStop}%
\bibitem [{\citenamefont {Genes}\ \emph {et~al.}(2008)\citenamefont {Genes},
  \citenamefont {Mari}, \citenamefont {Tombesi},\ and\ \citenamefont
  {Vitali}}]{ThreeEntOpto1}%
  \BibitemOpen
  \bibfield  {author} {\bibinfo {author} {\bibfnamefont {C.}~\bibnamefont
  {Genes}}, \bibinfo {author} {\bibfnamefont {A.}~\bibnamefont {Mari}},
  \bibinfo {author} {\bibfnamefont {P.}~\bibnamefont {Tombesi}}, \ and\
  \bibinfo {author} {\bibfnamefont {D.}~\bibnamefont {Vitali}},\ }\href
  {\doibase 10.1103/PhysRevA.78.032316} {\bibfield  {journal} {\bibinfo
  {journal} {Phys. Rev. A}\ }\textbf {\bibinfo {volume} {78}},\ \bibinfo
  {pages} {032316} (\bibinfo {year} {2008})}\BibitemShut {NoStop}%
\bibitem [{\citenamefont {Wang}\ \emph {et~al.}(2015)\citenamefont {Wang},
  \citenamefont {Chesi},\ and\ \citenamefont {Clerk}}]{ThreeEntOpto2}%
  \BibitemOpen
  \bibfield  {author} {\bibinfo {author} {\bibfnamefont {Y.-D.}\ \bibnamefont
  {Wang}}, \bibinfo {author} {\bibfnamefont {S.}~\bibnamefont {Chesi}}, \ and\
  \bibinfo {author} {\bibfnamefont {A.~A.}\ \bibnamefont {Clerk}},\ }\href
  {\doibase 10.1103/PhysRevA.91.013807} {\bibfield  {journal} {\bibinfo
  {journal} {Phys. Rev. A}\ }\textbf {\bibinfo {volume} {91}},\ \bibinfo
  {pages} {013807} (\bibinfo {year} {2015})}\BibitemShut {NoStop}%
\bibitem [{\citenamefont {Ma}\ \emph {et~al.}(2017)\citenamefont {Ma},
  \citenamefont {Miao}, \citenamefont {Pang}, \citenamefont {Evans},
  \citenamefont {Zhao}, \citenamefont {Harms}, \citenamefont {Schnabel},\ and\
  \citenamefont {Chen}}]{Ma17}%
  \BibitemOpen
  \bibfield  {author} {\bibinfo {author} {\bibfnamefont {Y.}~\bibnamefont
  {Ma}}, \bibinfo {author} {\bibfnamefont {H.}~\bibnamefont {Miao}}, \bibinfo
  {author} {\bibfnamefont {B.~H.}\ \bibnamefont {Pang}}, \bibinfo {author}
  {\bibfnamefont {M.}~\bibnamefont {Evans}}, \bibinfo {author} {\bibfnamefont
  {C.}~\bibnamefont {Zhao}}, \bibinfo {author} {\bibfnamefont {J.}~\bibnamefont
  {Harms}}, \bibinfo {author} {\bibfnamefont {R.}~\bibnamefont {Schnabel}}, \
  and\ \bibinfo {author} {\bibfnamefont {Y.}~\bibnamefont {Chen}},\ }\href@noop
  {} {\bibfield  {journal} {\bibinfo  {journal} {Nature Physics}\ }\textbf
  {\bibinfo {volume} {13}},\ \bibinfo {pages} {776 EP } (\bibinfo {year}
  {2017})}\BibitemShut {NoStop}%
\bibitem [{\citenamefont {Nimmrichter}\ \emph {et~al.}(2014)\citenamefont
  {Nimmrichter}, \citenamefont {Hornberger},\ and\ \citenamefont
  {Hammerer}}]{Nimmrichter14}%
  \BibitemOpen
  \bibfield  {author} {\bibinfo {author} {\bibfnamefont {S.}~\bibnamefont
  {Nimmrichter}}, \bibinfo {author} {\bibfnamefont {K.}~\bibnamefont
  {Hornberger}}, \ and\ \bibinfo {author} {\bibfnamefont {K.}~\bibnamefont
  {Hammerer}},\ }\href {\doibase 10.1103/PhysRevLett.113.020405} {\bibfield
  {journal} {\bibinfo  {journal} {Phys. Rev. Lett.}\ }\textbf {\bibinfo
  {volume} {113}},\ \bibinfo {pages} {020405} (\bibinfo {year}
  {2014})}\BibitemShut {NoStop}%
\bibitem [{\citenamefont {McMillen}\ \emph {et~al.}(2017)\citenamefont
  {McMillen}, \citenamefont {Brunelli}, \citenamefont {Carlesso}, \citenamefont
  {Bassi}, \citenamefont {Ulbricht}, \citenamefont {Paris},\ and\ \citenamefont
  {Paternostro}}]{Sam17}%
  \BibitemOpen
  \bibfield  {author} {\bibinfo {author} {\bibfnamefont {S.}~\bibnamefont
  {McMillen}}, \bibinfo {author} {\bibfnamefont {M.}~\bibnamefont {Brunelli}},
  \bibinfo {author} {\bibfnamefont {M.}~\bibnamefont {Carlesso}}, \bibinfo
  {author} {\bibfnamefont {A.}~\bibnamefont {Bassi}}, \bibinfo {author}
  {\bibfnamefont {H.}~\bibnamefont {Ulbricht}}, \bibinfo {author}
  {\bibfnamefont {M.~G.~A.}\ \bibnamefont {Paris}}, \ and\ \bibinfo {author}
  {\bibfnamefont {M.}~\bibnamefont {Paternostro}},\ }\href {\doibase
  10.1103/PhysRevA.95.012132} {\bibfield  {journal} {\bibinfo  {journal} {Phys.
  Rev. A}\ }\textbf {\bibinfo {volume} {95}},\ \bibinfo {pages} {012132}
  (\bibinfo {year} {2017})}\BibitemShut {NoStop}%
\bibitem [{\citenamefont {Mari}\ and\ \citenamefont {Eisert}(2009)}]{Mari2009}%
  \BibitemOpen
  \bibfield  {author} {\bibinfo {author} {\bibfnamefont {A.}~\bibnamefont
  {Mari}}\ and\ \bibinfo {author} {\bibfnamefont {J.}~\bibnamefont {Eisert}},\
  }\href {\doibase 10.1103/PhysRevLett.103.213603} {\bibfield  {journal}
  {\bibinfo  {journal} {Physical Review Letters}\ }\textbf {\bibinfo {volume}
  {103}},\ \bibinfo {pages} {213603} (\bibinfo {year} {2009})}\BibitemShut
  {NoStop}%
\end{thebibliography}%


\begin{widetext}
\clearpage

\setcounter{equation}{0}
\setcounter{figure}{0}
\setcounter{table}{0}
\setcounter{page}{1}
\makeatletter
\renewcommand{\theequation}{S\arabic{equation}}
\renewcommand{\thefigure}{S\arabic{figure}}

\begin{center}
	\textbf{\large Supplementary Material:\\ Conditional dynamics of optomechanical two-tone backaction-evading measurements}
\end{center}

\section{Details about the  optomechanical conditional dynamics}\label{AppA}
\subsection{Two-mode optomechanical system}
In the following we provide the explicit expressions of the terms appearing in Eq.~\eqref{Cond2}, necessary to quantify the conditional dynamics of the continuously monitored system, together with their derivation. We will employ the phase-space formalism, which is particularly convenient for our problem, and in particular we will follow closely the treatment of Ref.~\cite{AlessioBook}. We start by rewriting  the linearized optomechanical Hamiltonian Eq.~\eqref{Hint} in terms of the quadrature vector  $\hat {\rm x}=(\hat X_c,\hat P_c,\hat X_m, \hat P_m)^T$, which takes the form  $\hat H_I(t)=\frac12\hat {\rm x}^T S \hat {\rm x}$, with the matrix $S$ given by
\begin{equation}\label{HOM}
S=
\left(
\begin{array}{cccc}
0 & 0 & -g(1+\cos2\omega_m t) & - g\sin2\omega_m t\\[1ex]
0 & 0 & 0 & 0 \\[1ex]
-g(1+\cos2\omega_m t) & 0 & 0 & 0 \\[1ex]
- g\sin2\omega_m t& 0 & 0 & 0
\end{array}
\right).
\end{equation}
The system-bath coupling $\hat H_{\rm diss}$ is modeled by an energy-preserving interaction between each of the system modes and the excitations of two distinct baths, namely
\begin{equation}\label{Hdiss}
\hat H_{\rm diss}=i\sqrt\kappa (\hat a^\dag \hat \xi_c-\hat a \hat \xi_c^\dag)+
i\sqrt\gamma (\hat b^\dag \hat \xi_m-\hat b \hat \xi_m^\dag) \,,
\end{equation}
which is valid in the weak-coupling limit. For the mechanical system, the limit $\gamma_m\ll\omega_m$ is also understood, where the damping mechanism of quantum Brownian motion reduces to standard quantum-optical dissipation. The environmental modes $\hat \xi_{c,m}=\hat \xi_{c,m}(t)$ are labeled by time and provide a microscopic description of a white-noise process. In terms of quadrature operators the latter condition is expressed by $\langle \{\hat{\rm x}_b(t) ,\hat{\rm x}_b(t')\}\rangle=\sigma_b\delta(t-t')$, where we defined the vector $\nobreak{\hat{\rm x}_b(t)=(\hat X_{\xi_c}(t),\hat P_{\xi_c}(t),\hat X_{\xi_m}(t), \hat P_{\xi_m}(t))^T}$, each quadrature operator being defined analogously to system quadratures, and 
\begin{equation}
\sigma_b=\mathrm{diag}\left[\frac12,\frac12,\bar n+\frac12,\bar n+\frac12\right]\,,
\end{equation}
with $\bar n$ the thermal occupation of the mechanical bath. Similarly to the optomechanical coupling $H_I$, the bilinear interaction~\eqref{Hdiss} can be written as       
\begin{equation}\label{HC}
\hat H_{\rm diss}=\hat {\rm x}^T C \hat {\rm x}_b\, ,
\end{equation}
where the matrix $C$ is given by  
$C=\sqrt{\kappa}\,\boldsymbol{\omega}^{-1}\oplus\sqrt{\gamma}\,\boldsymbol{\omega}^{-1}$,
and we introduced the symplectic form $\boldsymbol{\omega}=
 \left(
\begin{array}{cc}
0 & 1\\
-1 &0
\end{array}\right)$.  
With these ingredients at hand, the  drift ($A$) and diffusion ($D$) matrices appearing in Eq.~\eqref{Cond2} can be expressed as~\cite{AlessioBook,Marco16} 
\begin{align}
A&=\bOmega S+\frac12\bOmega C\bOmega C^T\, ,\\
D&=\bOmega C \sigma_b C^T \bOmega^T\ ,
\end{align}
with $\bOmega=\boldsymbol{\omega}\oplus\boldsymbol{\omega}$. Their explicit expression reads
\begin{align}
A=\left(
\begin{array}{cccc}
 -\frac{\kappa }{2} & 0 & 0 & 0 \\[1ex]
 0 & -\frac{\kappa }{2} & g(1+\cos2\omega_m t) &  g\sin2\omega_m t \\[1ex]
 - g\sin2\omega_m t & 0 & -\frac{\gamma }{2} & 0 \\[1ex]
 g(1+\cos2\omega_m t) & 0 & 0 & -\frac{\gamma }{2} \\[1ex]
\end{array}
\right) \,,
\end{align}
\begin{equation}
D=\mathrm{diag}\left[\frac{\kappa}{2},\frac{\kappa}{2},\left(\bar n+\frac12\right)\gamma,\left(\bar n+\frac12\right)\gamma\right] \,.
\end{equation}
\par
We also need to incorporate the measurement process into the dynamical evolution. We consider the case of continuous monitoring of the output cavity field via homodyne detection. This measurement can be described by a projection onto a pure squeezed state, which is modeled by the following covariance matrix   
\begin{equation}\label{SigmaMeas1}
\sigma_{\mathrm{meas}}=\frac12R_{\theta}\,\textrm{diag}(r,r^{-1})\,R_{\theta}^T \, ,
\end{equation}
where $R_{\theta}$ is a rotation matrix. In particular, homodyne detection of the optical phase quadrature is recovered in the limit  $r\rightarrow 0$ and $\theta=\pi/2$. It is also desirable to account for non-unit efficiency of the detection process, which is modeled by a beam splitter of transmissivity $\sqrt \eta$ prior to the detection, and gives
\begin{equation}\label{SigmaMeas}
\sigma_{\mathrm{meas}}^{\eta}=\frac{1}{\eta}\sigma_\mathrm{meas}+\frac{1-\eta}{2\eta} \mathbb{1}\, .
\end{equation}
The matrices $B$ and $N$ describing the effect of the measurement on the environment in Eq.~\eqref{Cond2} are given by
\begin{equation}\label{MeasMatrices}
B=C\bOmega(\sigma_b+\sigma_{\mathrm{meas}}^{\eta})^{-\frac12}\, ,\qquad N=\bOmega C\sigma_b(\sigma_b+\sigma_{\mathrm{meas}}^{\eta})^{-\frac12}\, .
\end{equation}
We point out that, since we are interested in the case where only the photonic modes undergo monitoring, the correct way of evaluating Eq.~\eqref{MeasMatrices} is to take the covariance matrix of a bipartite measurement [i.e., Eq.~\eqref{SigmaMeas} for both optical and mechanical modes] and then taking the limit of vanishing efficiency on the mechanical modes, which corresponds to no monitoring of the mechanical environment. 

\par
\subsection{Three-mode optomechanical system}
In the case of a three-mode optomechanical system, the expressions entering the conditional evolution of the covariance matrix~\eqref{Cond2} can be easily deduced following the construction outlined above. In particular, the quadrature vector is now given by $\hat {\rm x}=(\hat X_c,\hat P_c,\hat X_{m,1}, \hat P_{m,1},\hat X_{m,2}, \hat P_{m,2})^T$ and the expression of the optomechanical interaction~\eqref{TwoToneQND} in terms of the quadratures reads
\begin{align}\label{HIThree}
\hat H_I(t)&=\frac{\Omega}{2}\sum_{j=1,2}(-1)^{j+1}(\hat X_{m,j}^2+\hat P_{m,j}^2)-g\sum_{j=1,2}\hat X_c[\hat X_{m,j}(1+\cos2\omega t)+\hat P_{m,j}\sin2\omega t]\,.
\end{align}
For simplicity, in our study we consider the case of equal single-photon optomechanical couplings, equal mechanical damping rates, and same occupancies of the baths. For non-degenerate mechanical modes, these conditions entail adjusting the local temperatures of the baths to achieve the same occupancy. However, we stress that our analysis can be easily extended to the case of asymmetric couplings and/or damping rates to describe experimental inaccuracies. The expressions entering Eq.~\eqref{Cond2} are given by
\begin{align}
A&=\left(
\begin{array}{cccccc}
 -\frac{\kappa }{2} & 0 & 0 & 0 & 0 & 0 \\[1ex]
 0 & -\frac{\kappa }{2} & g(1+\cos2\omega t) & g\sin2\omega t & g(1+\cos2\omega t) & g\sin2\omega t \\[1ex]
 -g\sin2\omega t & 0 & -\frac{\gamma }{2} & \Omega & 0 & 0\\[1ex]
  g(1+\cos2\omega t) & 0 & -\Omega & -\frac{\gamma }{2} & 0 & 0\\[1ex]
   -g\sin2\omega t & 0 & 0 & 0 & -\frac{\gamma }{2} & -\Omega \\[1ex]
  g(1+\cos2\omega t) & 0 & 0 & 0 & \Omega & -\frac{\gamma }{2} \\[1ex]
\end{array}
\right)\,, \\[1ex]
D&=\mathrm{diag}\left[\frac{\kappa}{2},\frac{\kappa}{2},\left(\bar n+\frac12\right)\gamma,\left(\bar n+\frac12\right)\gamma,\left(\bar n+\frac12\right)\gamma,\left(\bar n+\frac12\right)\gamma\right] \,, \\[1ex]
\sigma_b&=\mathrm{diag}\left[\frac12,\frac12,\bar n+\frac12,\bar n+\frac12,\bar n+\frac12,\bar n+\frac12\right] \,, \\[1ex]
C&=\sqrt{\kappa}\,\boldsymbol{\omega}^{-1}\oplus\sqrt{\gamma}\,\boldsymbol{\omega}^{-1}\oplus\sqrt{\gamma}\,\boldsymbol{\omega}^{-1} \,, \\[1ex]
\boldsymbol{\Omega}&=\boldsymbol{\omega}\oplus\boldsymbol{\omega}\oplus\boldsymbol{\omega}\,.
\end{align}

\section{expression of the steady-state conditional covariance matrix}
The matrix equation~\eqref{Cond2} can be solved exactly at the steady state. Besides the expression of the conditional mechanical variances $\sigma_{3,3}\equiv \sigma^2_{X_m}$ and
$\sigma_{4,4}\equiv \sigma^2_{P_m}$, respectively  given in Eq.~\eqref{VarQcond} and \eqref{VarPcond}, the other elements of the covariance matrix read
\begin{align}
&\sigma_{1,1}\equiv\sigma^2_{X_c}=\frac12\,,   &\sigma_{2,2}&\equiv\sigma^2_{P_c}=\frac{1}{4\eta \kappa} \left[ \sqrt{\gamma^2+\kappa^2+2\zeta} +\kappa (2\eta-1)-\gamma \right] \,,&  \\
&\sigma_{1,4}=\frac{g}{\gamma+\kappa}\,, &\sigma_{2,3}&=\frac{1}{8g\eta \kappa} \left[ \zeta +\gamma^2 -\gamma\sqrt{\gamma^2+\kappa^2+2\zeta}\right] \,,&
\end{align}
while $\sigma_{1,2}=\sigma_{1,3}=\sigma_{2,4}=\sigma_{3,4}=0$. We recall that we set $\zeta=\sqrt{\gamma\kappa[16 g^2\eta(1+2\bar n)+\gamma\kappa]}$. One can check that the optical phase quadrature is never squeezed, and that steady-state optomechanical entanglement can be present for suitable values of the parameters.  

\section{Perturbative solution for the effect of counterrotating terms}
In order to gain some analytical understanding of the long-time behavior of the covariance matrix associated to Eq.~\eqref{Hint}, we expand the latter in Fourier components~\cite{Mari2009} $\sigma(t)=\sum_n\exp(in2\omega_m t)\sigma_n$, retaining only the leading-order contribution $\sigma_{\pm1}$ for simplicity. In principle, truncating at sufficiently high order yields a set of algebraic equations that capture the steady-state covariance matrix, but as we already have a numerical method, we instead aim to obtain simple closed-form solutions and only perform second order perturbation theory in $H_{\mathrm{CR}}$. Given the solution in RWA $\sigma_{0}$ (see previous section), we find $\sigma_{1}$, which fulfills 
\begin{equation}
  0=-2 i\omega_m \sigma_1+A_0\sigma_{1}+\sigma_1A_0^T+A_1\sigma_0+\sigma_0A_1^T-\sigma_0BB^T\sigma_1-\sigma_1BB^T\sigma_0+\sigma_1BN+NB^T\sigma_1,
  \label{eq:sigma_1}
\end{equation}
where we have also introduced Fourier components of the coupling matrix $A(t)=\sum_n\exp(in2\omega_mt)A_{n}$. This equation is linear in $\sigma_1$, which means that it can readily be obtained from the RWA solution for $\sigma_0$ (note that $\sigma_{-1}=\sigma_1^\dagger$).
The $n\neq0$ Fourier components of the covariance matrix cause oscillating variances associated to a periodic steady state, which is the reason why in \cref{f:PlotSq} the squeezing corresponds to a shaded area rather than a single value. Physically, $A_{\pm1}$ in the above expression are a modulated coupling of the quadratures. To first order, they are a source term for the oscillating variances. The coupling between the quadratures is not QND, such that information about the previously unmonitored quadrature $P_m$ now enters the cavity via $A_{\pm1}$. The last four terms in Eq.~\eqref{eq:sigma_1} entail that, as a result of mixing of oscillating and stationary parts, the cavity output and thus the conditioning due to the measurement also oscillates. 

To second order in the counterrotating terms, they affect the stationary part of the covariance matrix as $\sigma_{0}+\sigma_{0,\mathrm{correction}}$, with the correction  given by
\begin{equation}
  \sigma_{0,\mathrm{correction}}=A_{-1}\sigma_1+\sigma_1A_{-1}^T+A_1\sigma_{-1}+\sigma_{-1}A_1^T-\sigma_{-1}BB^T\sigma_1-\sigma_1BB^T\sigma_{-1}.
  \label{eq:second_order_correction}
\end{equation}
Again we can distinguish two types of contributions. The terms containing $A_{\pm1}$ arise due to the unitary dynamics induced through the CR terms, whereas the terms containing $B$ are a result of the measurement. 
Deep in the backaction-dominated regime, the correction to the variance of the squeezed quadrature arises entirely from the dynamical part
and reads
\begin{equation}
  \sigma_{X_m,\mathrm{correction}}^2=\frac{\kappa}{2\omega_m}|\chi_c(2\omega_m)|^2g^2\frac{1}{2} + \mathcal O(\gamma/\omega_m)\,,
  \label{eq:Xm_correction}
\end{equation}
where $\chi_c(\omega)=(\kappa/2-i \omega)^{-1}$ is the cavity susceptibility. This contribution can be interpreted as measurement backaction (or shot noise) from the cavity entering the squeezed mechanical quadrature due to the CR terms. The fact that it results from cavity sidebands off resonance is captured by the cavity susceptibility evaluated at the position of the next-order sidebands at $2\omega_m$. On the other hand, the absence of the measurement efficiency clearly indicates that this is a dynamical effect. This is the dominant leading-order source of squeezing loss.

We can also look at the correction to the anti-squeezed quadrature, which to lowest order in $\gamma/\omega_m$ is
\begin{equation}
  \sigma_{P_m,\mathrm{correction}}^2=-\eta\frac{\kappa}{2\omega_m}|\chi_c(2\omega_m)|^2g^2\frac12(\mathcal C+2\bar n+1)^2+\mathcal O(\gamma/\omega_m),
  \label{eq:Pm_correction}
\end{equation}
where for convencience we have kept both $g^2$ and $\mathcal C$, which adds slight inconsistencies in the expansion for low $\gamma$. Comparison to the full second-order solution obtained from \cref{eq:second_order_correction} shows that \cref{eq:Pm_correction} is indeed a very good approximation. There is a striking similarity between the lowest-order correction to the mechanical quadratures \cref{eq:Xm_correction,eq:Pm_correction}, as both result from a coupling to the cavity sideband at $2\omega_m$. Interestingly, the correction to the anti-squeezed quadrature is negative, which means that the variance is decreased. Physically, the CR terms lead to some coupling of the anti-squeezed quadrature into the optical phase quadrature, such that the measurement reduces the uncertainty in $P_m$. This conclusion is supported by the fact that the whole expression is proportional to the measurement efficiency. As this reduction is larger in magnitude than the correction to the variance of $X_m$, the mechanical state overall is purified, a conclusion that is borne out by our numerical simulations. 

\end{widetext}
\end{document}